\documentclass{article}
\usepackage[utf8]{inputenc}
\usepackage[margin=1.5in]{geometry}
\usepackage{amsmath}
\usepackage{amsfonts}
\usepackage{amsmath}
\usepackage{amssymb}
\usepackage{mathrsfs}
\usepackage{amsthm}
\usepackage{mathtools}
\usepackage{natbib}
\usepackage[pdftex]{graphicx}
\usepackage{listings}
\usepackage{color}
 
\definecolor{codegreen}{rgb}{0,0.6,0}
\definecolor{codegray}{rgb}{0.5,0.5,0.5}
\definecolor{codepurple}{rgb}{0.58,0,0.82}
\definecolor{backcolour}{rgb}{0.95,0.95,0.92}
 
\lstdefinestyle{mystyle}{
    basicstyle=\tiny,
    backgroundcolor=\color{backcolour},   
    commentstyle=\color{codegreen},
    keywordstyle=\color{magenta},
    numberstyle=\tiny\color{codegray},
    stringstyle=\color{codepurple},
    basicstyle=\footnotesize,
    breakatwhitespace=false,         
    breaklines=true,                 
    captionpos=b,                    
    keepspaces=true,                 
    numbers=left,                    
    numbersep=5pt,                  
    showspaces=false,                
    showstringspaces=false,
    showtabs=false,                  
    tabsize=2
}
 
\newcommand{\gau}{\mathcal{N}}

\newcommand{\be}{\begin{equation}}
\newcommand{\ee}{\end{equation}}

\newcommand{\GP}{ \mathcal{GP}}

\newcommand{\mathd}{\mathrm{d}}

\newcommand{\bm}{{\bf m}}

\newcommand{\xv}{\mathbf{x}}
\newcommand{\Xv}{\mathbf{X}}
\newcommand{\muv}{\mathbf{\mu}}
\newcommand{\yv}{\mathbf{y}}

\newcommand{\fv}{\mathbf{f}}

\newcommand{\mbf}[1]{{\boldsymbol{\mathbf{#1}}}}
\renewcommand{\bm}{\mbf}

\usepackage{color}

\DeclareMathOperator*{\argmax}{arg\,max}

\newcommand{\op}[1]{\ensuremath{\operatorname{#1}}}
\title{Gaussian Process Regression for Derivative Portfolio Modeling and Application to CVA Computations}
 \author{St\'{e}phane Cr\'{e}pey\thanks{St\'{e}phane Cr\'{e}pey is a Professor in the Department of Mathematics, University of Evry,  Paris Saclay. E-mail: stephane.crepey@univ-evry.fr. The research of S. Cr\'{e}pey benefited from the support of the Chair Stress Test, RISK Management and Financial Steering, led by the French Ecole polytechnique and its Foundation and sponsored by BNP Paribas.}\\
 LaMME, Univ Evry, CNRS, Universit\'e Paris-Saclay, 91037, Evry, France\\
    \\
    and\\
    \\
    Matthew F. Dixon\thanks{Matthew Dixon is an Assistant Professor in the Department of Applied Mathematics, Illinois Institute of Technology, Chicago. E-mail: matthew.dixon@iit.edu. The research of M. Dixon is supported by a grant from Intel Corp.\newline  
    \textit{Acknowledgement:} The authors are grateful to Marc Chataigner, Areski Cousin, Mike Ludkovski and the anonymous referees, for insights and feedback, and to Bouazza Saadeddine for  the generation of the mark-to-market cube that served as a basis for the regression exercise of Section \protect\ref{ss:scalcube}. An earlier version of this paper was presented at Quantminds 2019 in Vienna, SIAM FME 2019 in Toronto, and AMAMEF 2019 in Paris. }\\
    Department of Applied Mathematics,
    Illinois Institute of Technology. 
    }

\def\qqq{\quad\quad\quad}

\newcommand{\beql}[1]{\begin{eqnarray}\label{#1}\begin{aligned}}
\newcommand{\eeql}{\end{aligned}\end{eqnarray}}

\newcommand{\bel}{\begin{eqnarray*}\begin{aligned}}
\newcommand{\eel}{\end{aligned}\end{eqnarray*}}

\def\eee{\end{document}}

\def\mean\mathbb{E}\def\mean{\textrm{mu}}\def\mean{\mathrm{mean}}\def\mean{\textrm{E}}\def\mean{E}

\def\citep{\citet}
\begin{document}

\maketitle
 
\begin{abstract}
Modeling counterparty risk is computationally challenging because
it requires the simultaneous evaluation of all the trades with  each  counterparty under both market and credit risk. We present a multi-Gaussian process regression approach, which is well suited for OTC derivative portfolio valuation involved in CVA computation. Our approach avoids nested simulation or simulation and regression of cash flows by learning a Gaussian metamodel for the mark-to-market cube of a derivative portfolio. We model the joint posterior of the derivatives as a Gaussian process over function space,  with the spatial covariance structure imposed on the risk factors. Monte-Carlo simulation is then used to simulate the dynamics of the risk factors. The uncertainty in portfolio valuation arising from the Gaussian process approximation is quantified numerically. Numerical experiments demonstrate the accuracy and convergence properties of our approach for CVA computations, including a counterparty portfolio of interest rate swaps.\end{abstract}

\def\keywordname{{\bfseries Keywords:}}
\def\keywords#1{\par\addvspace\baselineskip\noindent\keywordname\enspace
\ignorespaces#1}
 
\vspace{2mm}



\section{Introduction}
Post the global financial crisis of 2007-2008, banks have been subject to much stricter regulation and conservative capital and liquidity requirements. Pricing, valuing and managing over-the-counter (OTC) derivatives has been substantially revised to more robustly capture counterparty credit risk.  Pricing and accounting  now includes  valuation  adjustments  collectively  known  as  XVAs (\citep{abbasturki:hal-01714747, RePEc:arx:papers:1412.5332, crepey2014}). 
The BCBS pointed out that 2/3 of total credit losses during the 2007-2009 crisis were CVA losses, i.e.~CVA increases, where the CVA liability of a bank is its expected loss triggered by future counterparty defaults.
As a consequence, a CVA capital charge has been introduced since the initial phase of the Basel III framework in December 2010.  

Modeling counterparty risk is computationally challenging because
it requires the evaluation of all the trades
with  each  counterparty under market and credit simulation.  For instance, CVA computation requires pathwise pricing of each counterparty portfolio under simulated market moves, with counterparty default modeled separately. The sensitivities of the CVA, with respect to all the underlying market risk buckets, are required for hedging.  

The main source of computational complexity in XVA computations arises from the necessity of revaluing portfolio holdings (including path dependent or early exercise options) in numerous future dynamic scenarios. 
In the case of XVA (first-order) sensitivities,
there has been much progress towards real-time estimation using adjoint algorithmic differentiation (\citep{giles2005, Capriotti2011a, Capriotti2011b, AntonovIssakovMcClellandMechkov2018,HugeSavine17}). However, algorithmic differentiation is still very challenging to implement at the level of a banking derivative portfolio, and it typically comes at the cost of more or less drastic simplifications of the xVA metrics to be differentiated. Hence, bump-and-revalue sensitivities remain useful (and are in fact unavoidable regarding second-order sensitivities) and, again, multiple and fast valuation is required.

In this paper, we 
investigate the possible use of
Gaussian processes (GP) regression as a metamodel of the mark-to-market (MtM) cube, i.e.~the value of client portfolios in future time points and scenarios. 
Our approach consists in simulating the market risk factors forward in time and then interpolating the mark-to-market cube
from a set of model generated, reference derivative prices. Such an approach is predicated on the notion that a GP model, once trained,
 can provide fast and reliable prices (as well as the associated, analytically differentiated Greeks).

 We refer the reader to \citep{Rasmussen:2005:GPM:1162254} for a general introduction to Gaussian process regression, or simply Gaussian processes (GPs).
 As opposed to frequentist machine learning techniques, including neural networks or support vector machines, which only provide point estimates,
 GPs quantify the
uncertainty of their predictions\footnote{Through out this paper, we will refer to 'prediction' as out-of-sample point estimation. For avoidance of doubt, the test point need not be in the future as the terminology suggests.}. A high uncertainty in a prediction might result in a GP model estimate being rejected in favor of either retraining the model or even using full model revaluation. Another motivation for using GPs is the availability of  efficient training method for the model hyper-parameters.
In addition to a number of favorable statistical and mathematical properties, such as universality (see \citep{Micchelli:2006:UK}), the implementation support infrastructure is mature and provided by
open source machine learning packages such as \verb|GpyTorch|, \verb|scikit-learn|, \verb|Edward|, or \verb|STAN|.  

 
GPs have demonstrated much success in applications outside of finance and sometimes under the name of kriging.
The basic theory of prediction with Gaussian processes dates back to at least as far as the time series work of Kolmogorov or Wiener in the 1940s (see \citep{10.5749/j.ctttsphx}). 
Examples of applying GPs to financial time series prediction are presented in \citep{Roberts20110550}. These authors helpfully note that AR$(p)$ processes are discrete time equivalents of GP models with a certain class of covariance functions, known as Mat\'ern covariance functions. Hence, GPs can be viewed as a Bayesian non-parametric generalization of well known econometrics techniques. 
\citep{doi:10.1080/00036846.2016.1167827} 
present a GP method for optimizing financial asset portfolios.

The adoption of kriging methods in financial derivative modeling is more recent. The underlying data to which the GP is fitted are then typically generated by the user itself in a model,
rather than market data---somewhat counter the motivation for adopting machine learning, but also the case in other recent computational finance applications such as \citep{Hernandez17},
\citep{EHanJentzen17}, or \citep{BuhlerGononTeichmannWood18}.
The motivation is then fast pricing, once the prediction algorithm has been trained off-line as a pre-processing stage.

\citep{CousinMaatoukRulliere16} introduce shape-constrained GPs 
to ensure non-arbitrable and error-controlled yield-curve and CDS curve interpolation. 
\citep{LudkovskiGramacy15} and \citep{RePEc:arx:papers:1509.02179}) reformulate the Bermudan option pricing problem as a
response surface metamodeling problem, which they address by kriging.
In the context of expected  shortfall computations,
\citep{Liu_stochastickriging} and \citep{RePEc:arx:papers:1710.05204} use GPs to infer portfolio values in a given scenario, based on inner-level simulation of nearby scenarios.  This
significantly reduces the required computational effort by restricting inner-level simulations to few selected scenarios, while naturally taking account of the variance that arises from inner-level simulation.

\citep{Spiegeleer2018} propose offline learning of a derivative pricing function through Gaussian process regression. Specifically, the authors configure the training set over a grid and then use the GP to interpolate at the test points. 
They demonstrate the speed up of GPs relative to Monte-Carlo methods and tolerable accuracy loss applied to pricing and Greek estimation with a Heston model, in addition to approximating the implied volatility surface. 
The increased expressibility of GPs compared to cubic spline interpolation, a popular numerical approximation techniques useful for fast point estimation, is also demonstrated. 

However, the applications shown in \citep{Spiegeleer2018} are limited to single instrument pricing, they do not consider the portfolio aspects. In particular, their study is limited to single-response GPs, as opposed to also multi-response GPs in this work (respectively referred to as single- vs. multi-GPs for brevity).
 In a single-GP setting, individual GPs are used to model the posterior of each predicted derivative price under the assumption that the derivative prices are independent, conditional on the training data and test input. Given that either the derivatives may share common underlyings, or the underlyings are different but correlated, this assumption is clearly violated in practice.        
By contrast, multi-GPs (see \citep{AlvarezRosascoLawrence12} for a survey) directly model the uncertainty in the prediction of a vector of derivative prices (responses) with spatial covariance matrices specified by kernel functions. Thus the amount of {\it error} in the 
mark-to-market cube prediction (the prediction itself does not change)
can only be adequately modeled using multi-GPs.

\paragraph{Outline}  This paper uses single- and multi-GPs for learning the posterior distribution of a mark-to-market cube,
which is then used in the context of CVA computations.
Section \ref{s:gauss} reviews single-response GPs and Section \ref{s:single} illustrates their use for derivative pricing and Greeking applications. Section \ref{sect:multiGP} extends the setup to a multi-response generalization of GPs.
Section \ref{sect:CVA} deals with CVA computations using a
Monte Carlo GPs approach, whereby the GP predicted MtM cube is used for valuing the derivative portfolio of the bank at the nodes of a Monte Carlo simulation for the bank CVA.
The concluding Section \ref{sect:conclusion} summarizes our findings and puts GPs in perspective with either simpler or more elaborate regression alternatives. Some of the numerical examples are illustrated with Python code excerpts
demonstrating the key features of our
approach. All performances are based on a 2.2 GHz Intel Core i7 laptop.  These and additional examples are provided in the Github repository  \verb|https://github.com/mfrdixon/GP-CVA|.
The examples can be run using the command \verb|ipython notebook| (once the required packages have been loaded).\\

Note that our setup involves both
the randomness of financial risk factors and the Bayesian uncertainty relative to GP estimation. 
For clarity of exposition, we denote by $\mathbb{P}$ and $\mathbb{E}$ the probability and expectation with respect to a pricing measure, and by $\mean$ (respectively $\textrm{var}$ or $\textrm{cov}$) a GP point (respectively variance or covariance) estimate. A confidence interval refers to a Monte Carlo estimate relative to the randomness of the financial risk factors, whereas an uncertainty band refers to the GP estimation procedure (both computed at the 95\% probability level).

\section{Single-Output Gaussian Processes}\label{s:gauss}

This section is a primer on (standard, single-) Gaussian processes inference, written in the classical Bayesian statistics style. Financial readers not acquainted with it may refer to \citep{Rasmussen:2005:GPM:1162254}
\citep{MacKay97gaussianprocesses},
and 
\citep[Chapter 15]{Murphy12}
for more background and detail.

Statistical inference involves learning a function $Y=f(X)$ of the data, $(X,Y):=\{(\xv_i, y_i)~|~i=1,\dots,n\}$. 
The idea of Gaussian processes (GPs) is to, without parameterizing\footnote{This is in contrast to nonlinear regressions commonly used in finance, which attempt to parameterize a non-linear function with a set of weights.} $f(X)$, place a probabilistic prior directly on the space of functions. The GP is hence a Bayesian nonparametric model that generalizes the Gaussian distributions from finite dimensional vector spaces to infinite dimensional function spaces. 
GPs are an example of a more general class of supervised machine learning techniques referred to as `kernel learning', which model the covariance matrix from a set of parametrized kernels over the input. GPs extend and put in a Bayesian framework spline or kernel interpolators, as well as Tikhonov regularization (see \citep{Rasmussen:2005:GPM:1162254} and \citep{AlvarezRosascoLawrence12}). 
\citep{neal2012bayesian} also observed that
certain neural networks with one hidden layer converge to a Gaussian process in the limit of an infinite number of hidden units.

In this section we restrict ourselves to the simpler case of single-response GPs where $f$ is real-valued (multi-response GPs will be considered in Section \ref{sect:multiGP}).


\subsection{Gaussian Processes Regression and Prediction}\label{ss:gpinf}

We say that a random function $f:\mathbb{R}^p\mapsto \mathbb{R}$ is drawn from a GP with a mean function $\mu$ and a covariance function, called kernel, $k$, i.e.~$f \sim \GP(\mu, k)$, if for any input points $\xv_1, \xv_2, \dots, \xv_n$ in $\mathbb{R}^p$, the corresponding vector of function values is Gaussian:
\begin{equation*}
  [f(\xv_1), f(\xv_2), \dots, f(\xv_n)] \sim \gau(\muv, K_{X,X}),
\end{equation*}
for some mean vector $\muv$, such that $\muv_i = \mu(\xv_i)$, and covariance matrix $K_{X,X}$ that satisfies $(K_{X,X})_{ij} = k(\xv_i, \xv_j)$. Unless specified otherwise, we follow the convention\footnote{This choice is not a real limitation in practice (since it only regards the prior and it does not prevent the mean of the
predictor from being nonzero).} in the literature of assuming $\bm{\mu} = \bm{0}$.

Kernels $k$ can be any symmetric positive semidefinite function, which is
the infinite-dimensional analogue of the notion of a symmetric positive semidefinite (i.e.~covariance) matrix, i.e.~such that
$$\sum_{i,j=1}^n k(\xv_i,\xv_j)\xi_i\xi_j\ge 0\mbox{, for any points $\xv_k\in\mathbb{R}^p$ and reals $\xi_k$}.$$
Radial basis functions (RBF) are kernels that only depend on $||\xv-\xv'||$, such as the squared exponential
 (SE) kernel 
\beql{e:se}
&
k(\xv,\xv')=\exp \{-\frac{1}{2\ell^2}||\xv-\xv'||^2\},
\eeql
where the length-scale parameter $\ell$ can be interpreted as ``how far you need
to move in input space for the function values to become uncorrelated'',
or the Matern (MA) kernel 
\beql{e:mat} & 
k(\xv,\xv')=
{
\frac {2^{1-\nu }}{\Gamma (\nu )}}{\Bigg (}{\sqrt {2\nu }}{\frac {||\xv-\xv'||}{\ell}}{\Bigg )}^{\nu }K_{\nu }{\Bigg (}{\sqrt {2\nu }}{\frac {||\xv-\xv'||}{\ell }}{\Bigg )}  \eeql
(which converges to \eqref{e:se} in the limit where $\nu$ goes to infinity),
where $\ell$ and $\nu$ are non-negative parameters, $\Gamma$ is the gamma function, and $K_{\nu }$ is the modified Bessel function of the second kind.
One advantage of GPs over interpolation methods is their expressability. In particular, one can combine the kernels by convolution (cf. \citep{Melkumyan2011}). Moreover, the regularity of the GP interpolation is controllable through the one of the kernel.

GPs can be seen as distributions over the reproducing kernel Hilbert space (RKHS) of functions which is uniquely defined by the kernel function, $k$~(see \citep{Scholkopf:2001:LKS:559923}).
GPs with RBF kernels are known to be universal approximators with prior support to within an arbitrarily small epsilon band of any continuous function~(see \citep{Micchelli:2006:UK}).

GPs also provide ``differential regularity'' --- GPs are RKHSs defined in terms of differential operators, with the Hilbert norm of the latent function having the effect of penalizing the gradients. Regularity of the GP interpolation is thus controllable through the choice of the kernel and smoothing parameters (see Section 6.2 of \cite{Rasmussen:2005:GPM:1162254}).

One limitation of the kernel is that it does not reveal any hidden representations --- failing to identify the useful features for solving a particular problem. The issue of feature
discovery can be addressed by GPs through imposing ``spike-and-slab'' mixture priors on the covariance parameters (see \citep{savitsky2011}).

Assuming additive Gaussian i.i.d. noise, $y \mid \xv \sim \gau(f(\xv), \sigma^2)$, and a GP prior on $f(\xv)$, given training  inputs $\xv \in X$ and training targets $y\in Y$, the predictive distribution of the GP evaluated at arbitrary test points $X_*$ is:
\begin{equation}
  f_* \mid X, Y, X_* \sim \gau(\mean[f_* | X,  Y, X_*], \mathrm{var}[f_* | X, Y, X_*]),
\end{equation}
where the moments of the posterior over $X_*$ are
\begin{equation} \label{eq:posterior_moments}
  \begin{aligned}
    \mean[f_* | X, Y, X_*] &= \muv_{X_*} + K_{X_*,X} [K_{X,X} + \sigma^2 I]^{-1}Y,\\
    \mathrm{var}[f_* | X, Y, X_*] &= K_{X_*,X_*} - K_{X_*,X} [K_{X,X} + \sigma^2 I]^{-1} K_{X,X_*}.
  \end{aligned}
\end{equation}
Here, $K_{X_*,X}$, $K_{X,X_*}$, $K_{X,X}$, and $K_{X_*,X_*}$ are matrices that consist of the kernel, $k:\mathbb{R}^p\times \mathbb{R}^p\mapsto \mathbb{R}$, evaluated at the corresponding points, $X$ and $X_*$, and $\muv_{X_*}$ is the mean function evaluated on the test inputs $X_*$.

In the context of derivative pricing applications, $X$ may correspond to a set of risk factor grid nodes, $Y$ to the corresponding model prices (valued by analytical formulas or any, possibly approximate, classical numerical finance pricing schemes), $\mean[f_* | X, Y, \xv_*]$ to the GP regressed prices corresponding to the new value $\xv_*\in X_*$ of the risk factors, and $\mathrm{var}[f_* | X, Y, \xv_*] $ to the corresponding interpolation uncertainty. Note that the latter is only equal to 0 if $\xv_*\in X$ and one is in the noise-free case where $\sigma$ has been set to 0.

 We emphasize that, in a least square Monte-Carlo regression approach a la  \citep{doi:10.1093/rfs/14.1.113} (see e.g. \citep[Part IV]{cre}), we train function approximators, usually as linear combinations of fixed basis functions, on simulated samples. By contrast, GPs are trained on values (not samples), on a (structured or not, deterministically or stochastically generated) grid, like a sophisticated interpolator.

\subsection{Hyper-parameter Tuning}
GPs are fit to the data by optimizing \emph{the evidence}-the marginal probability of the data given the model with respect to the learned kernel hyperparameters.  

The evidence has the form (see e.g. \citep[Section 15.2.4, p. 523]{Murphy12}):
\begin{equation}\label{e:maxlik}
    \log p(Y \mid X, \lambda) = - \left[Y^\top (K_{X,X} + \sigma^2 I)^{-1}Y + \log \det (K_{X,X} + \sigma^2 I)\right]  - \frac{n}{2}\log 2\pi,
\end{equation}
where the kernel hyperparameters $\lambda$ include $\sigma$ in \eqref{e:maxlik} and parameters of
$K_{X,X}$  
(e.g. $\lambda=[\ell, \sigma]$, assuming an SE
kernel as per \eqref{e:se} or an MA kernel for some exogenously fixed value of $\nu$ in \eqref{e:mat}).

The first and second term in the brackets in \eqref{e:maxlik} can be interpreted as a \emph{model fit} and a \emph{complexity penalty} term (see 
\citep[Section 5.4.1]{Rasmussen:2005:GPM:1162254}).
Maximizing the evidence with respect to the kernel hyperparameters, i.e.~computing
$\lambda^*= \argmax_{\lambda} \log p(Y \mid X, \lambda) $,
results in an automatic Occam's razor 
controlling the trade-off between the regression fit and the regularity of the interpolator (see \citep[Section 2.3]{AlvarezRosascoLawrence12} and \citep{rasmussen2001occam}).
In practice, the negative evidence is minimized by stochastic gradient descent (SGD).
The gradient of the evidence is given analytically by 
\begin{equation}\label{e:aloglik}
\partial_{\lambda}  \log p(Y \mid X, \lambda)= tr\left(\mathbf{\alpha}\mathbf{\alpha}^T-(K_{X,X} + \sigma^2 I)^{-1}\right) \partial_{\lambda} (K_{X,X} + \sigma^2 I)^{-1},
\end{equation}
where $\mathbf{\alpha}:=(K_{X,X} + \sigma^2 I)^{-1}Y$, $\partial_{\sigma} (K_{X,X} + \sigma^2 I)^{-1}= -2\sigma(K_{X,X} + \sigma^2 I)^{-2},$ and, in the case of the SE or MA kernels,
\begin{eqnarray}
\partial_{\ell} (K_{X,X} + \sigma^2 I)^{-1}&=& -(K_{X,X} + \sigma^2 I)^{-2} \partial_{\ell} K_{X,X}
\end{eqnarray}
(with, in the SE case,
$
\partial_{\ell} k(\xv,\xv')=\ell^{-3}||\xv-\xv'||^2k(\xv, \xv')$).

\subsection{Computational Properties} \label{ss:comp}

Training time, required for maximizing \eqref{e:maxlik} numerically, scales poorly with the number of observations $n$.
This stems from the need to solve linear systems and
compute log determinants involving an $n\times n$ symmetric  positive  definite
covariance  matrix $K$. This  task
is  commonly  performed  by  computing  the  Cholesky
decomposition of $K$ incurring $\mathcal{O}(n^3)$ complexity. Prediction, however, is faster and can be performed in $\mathcal{O}(n^2)$ with a matrix-vector multiplication for each test point, and hence the primary motivation for using GPs is real-time risk estimation performance. 

If uniform grids are use, we have $n=\prod_{k=1}^{p} n_k$, where $n_k$ are the number of grid points per variable. However, mesh-free GPs can be used as described in Section \ref{sect:mesh-free}.

In terms of storage cost, although each kernel matrix $K_{X,X}$ is $n \times n$, we only store the n-vector $\alpha$ in \eqref{e:aloglik}, which brings reduced memory requirements.

\paragraph{Massively scalable Gaussian processes} Massively scalable Gaussian processes (MSGP) are a recent significant extension of the basic kernel interpolation framework described above. The core idea of the framework, which 
is detailed in~\citep{gardner2018product},
is to improve scalability by combining GPs with `inducing point methods'. Using structured kernel interpolation (SKI), a small set of $m$ inducing points are carefully selected from the original training points. Under certain choices of the kernel, such as RBFs, a Kronecker and Toeplitz structure of the covariance matrix can be exploited by fast Fourier transform (FFT). Finally, output over the original input points is interpolated from the output at the inducing points. The interpolation complexity scales linearly with dimensionality $p$ of the input data by expressing the kernel interpolation as a product of 1D kernels. Overall, SKI gives $\mathcal{O}(pn+ pmlog m)$ training complexity and $\mathcal{O}(1)$ prediction time per test point, using the LanczOs  
Variance Estimates of \citep{DBLP:journals/corr/abs-1803-06058}. 
In this paper, we primarily use the basic interpolation approach for simplicity. 

\paragraph{Online learning}
If the option pricing model is recalibrated intra-day, then the corresponding GP model should be retrained. Online learning techniques permit performing this incrementally (see \citep{4721437}). To enable online learning, the training data should be augmented with the constant model parameters. Each time the parameters are updated, a new observation $(\xv', y')$ is generated from the option model prices under the new parameterization. The posterior at test point $\xv_*$ is then updated with the new training point following

\be
p(f_* | X, Y,  \xv', y', \xv_* )= \frac{p(\xv', y' | f_*)p(f_* | X, Y, \xv_*)}{\int_{Z} p(\xv', y' | z)p(z | X, Y, \xv_*)dz},
\ee
where the previous posterior $p(f_* | X, Y, \xv_*)$ becomes the prior in the update and $f_*\in Z \subset \mathbb{R}$.
Hence the GP learns over time as model parameters (which are an input to the GP) are updated through pricing model recalibration.

\section{Pricing and Greeking With Single-Response Gaussian Processes}\label{s:single}

\subsection{Pricing}\label{sect:num_exp}

In the following example, a portfolio holds a long position in both a European call and a put option struck on the same underlying, with $K=100$. We assume that the underlying follows Heston dynamics (in risk-neutral form):

\beql{eq:heston}
\frac{dS_t}{S_t} &=& r dt  + \sqrt{V_t}dW_t^{1},\\
dV_t &=& \kappa(\theta -V_t)dt  + \sigma \sqrt{V_t}dW_t^{2},\\
 d\langle W^{1},   W^{2}\rangle_t&=&\rho dt,
\eeql
where the notation is defined in Table \ref{tab:Heston_params}. We use a Fourier Cosine method by \citep{Fang08anovel} to generate the European Heston option price training and testing data for the GP. We also use this method to compare the GP Greeks, obtained by differentiating the kernel function.
 
Table \ref{tab:Heston_params} also lists the values of the Heston parameters and terms of the European call and put option contract used in our numerical experiments. Additionally, the data is generated using an Euler time stepper for \eqref{eq:heston} using 100 time steps over a two year horizon.
\begin{table}[h!]
\centering
\begin{tabular}{|l|cc|}
\hline
Parameter description & Symbol & Value \\
\hline
Initial stock price & $S_0$ & 100 \\
Initial variance & $V_0$ & 0.1 \\
Mean reversion rate & $\kappa$ & 0.1\\
Mean reversion level & $\theta$ & 0.15\\
Vol. of Vol. & $\sigma$ & 0.1 \\
Risk free rate & $r$ & 0.01 \\ 
Strike & $K$ & 100\\
Maturity & $T$ & 2.0 \\
Correlation & $\rho$ & $-0.9$\\
\hline
\end{tabular}
\caption{This table shows the values of the parameters for the Heston dynamics and terms of the European call and put option contracts.}
\label{tab:Heston_params}
\end{table}

For each $t_i$ in a grid of dates (which in the context of CVA would correspond to the MtM exposure simulation times $t_i$, see Section \ref{sect:CVA}), we simultaneously fit numerous 
GPs to both gridded call and put prices over stock price $S$ and volatility $\sqrt{V}$ (keeping time to maturity fixed). Then we fit a GP to the Heston pricing function from Heston prices computed by Fourier formulas for the gridded
values of $S$ and $\sqrt{V}$. We emphasize that the Heston dynamics \eqref{eq:heston} are not used in the simulation mode in this procedure. 

\begin{figure}[h!]
\centering
\begin{tabular}{ccc}
(a) Price: $T-t=1.8$ & (b) Price: $T-t=1.0$ & (c) Price: $T-t=0.2$\\
\includegraphics[width=0.33\linewidth]{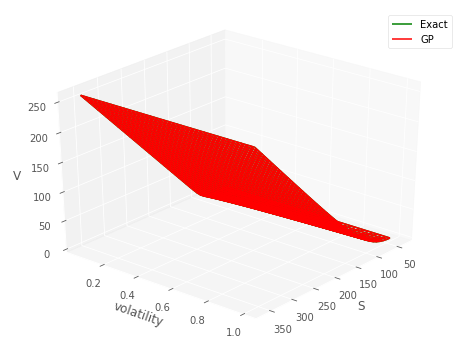} &
\includegraphics[width=0.33\linewidth]{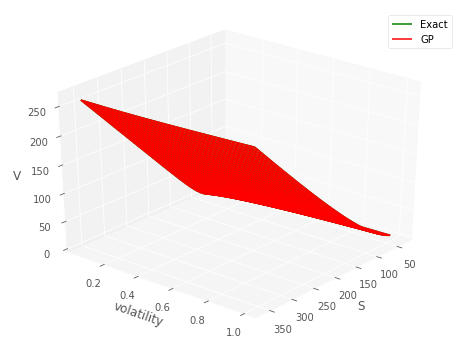} &
\includegraphics[width=0.33\linewidth]{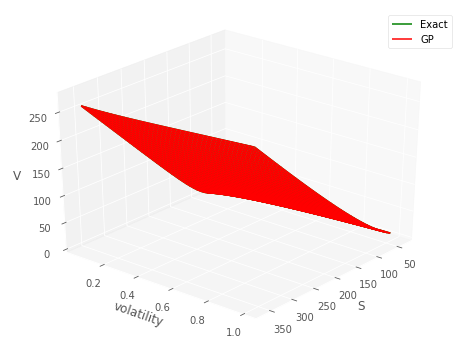}\\
(a) Error: $T-t=1.8$ & (b) Error: $T-t=1.0$ & (c) Error: $T-t=0.2$\\
\includegraphics[width=0.33\linewidth]{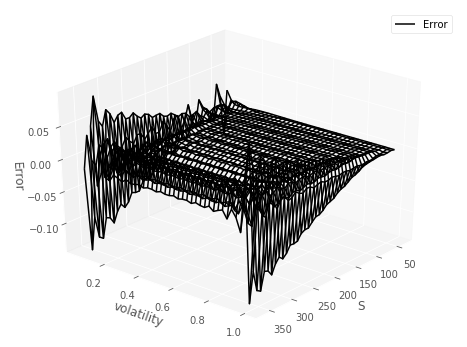} &
\includegraphics[width=0.33\linewidth]{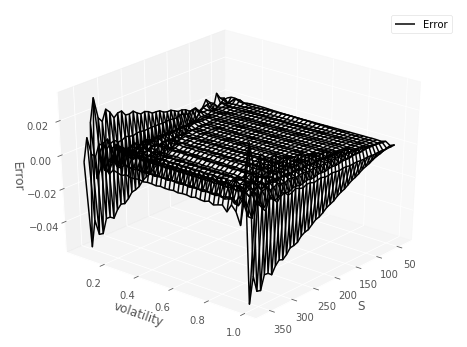} &
\includegraphics[width=0.33\linewidth]{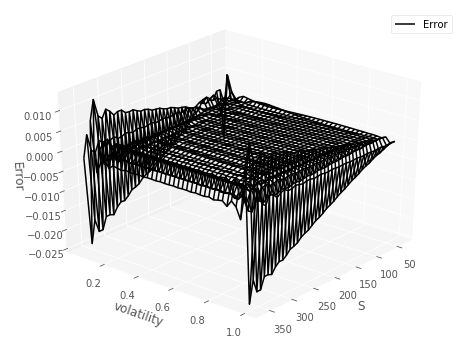}\\

\end{tabular}
\caption{\textit{This figure compares the gridded Heston GP and semi-analytic ('exact') model call prices (top) and error (bottom) surfaces at various time to maturities. The GP estimate is observed to be practically identical (on average it's slightly above the semi-analytic solution). Within each column in the figure, the same GP model has been simultaneously fitted to both the Heston model call and put price surfaces over a $30\times 30$ grid of prices and volatilities, fixing the time to maturity. Across each column, corresponding to different time to maturities, a different GP model has been fitted. The GP is then evaluated out-of-sample over a $40\times 40$ grid, so that many of the test samples are new to the model. This is repeated over various time to maturities.}}
\label{fig:GPS_surface_call}
\end{figure}

Listing \ref{list:Heston} details how the GP and data are prepared to predict prices over the two dimensional grid, for a fixed time to maturity and strike. Figures \ref{fig:GPS_surface_call} and \ref{fig:GPS_surface_call} (top) show the comparison between the gridded semi-analytic and GP call and put price surfaces at various time to maturities, together with the GP estimate. Within each column in the figures, the same GP model has been simultaneously fitted to both the call and put price surfaces over a $30\times 30$ grid $\Omega_h\subset \Omega:=[0,1]\times [0,1]$ of stock prices and volatilities\footnote{Note that the plot uses the original coordinates and not the re-scaled co-ordinates.}, for a given time to maturity. The bottom panel of the figure shows the error surfaces between the GP and semi-analytic estimates. The scaling to the unit domain is not essential. However, we observed superior numerical stability when scaling. 

Across each column, corresponding to different time to maturities, a different GP model has been fitted. The GP is then evaluated out-of-sample over a $40\times 40$ grid $\Omega_{h'}\subset \Omega$, so that many of the test samples are new to the model. This is repeated over the various dates $t_i$. The option model versus GP model are observed to produce very similar values.


\begin{figure}[h!]
\centering
\begin{tabular}{ccc}
(a) Price: $T-t=1.8$ & (b) Price: $T-t=1.0$ & (c) Price: $T-t=0.2$\\
\includegraphics[width=0.33\linewidth]{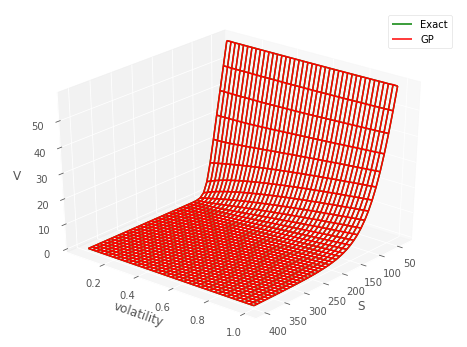} &
\includegraphics[width=0.33\linewidth]{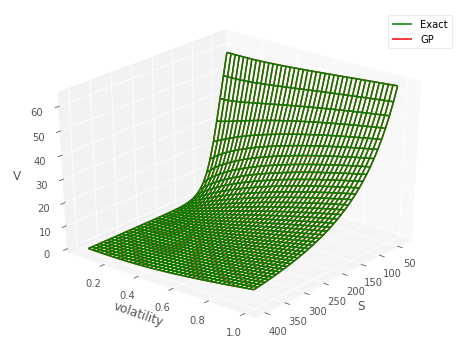} &
\includegraphics[width=0.33\linewidth]{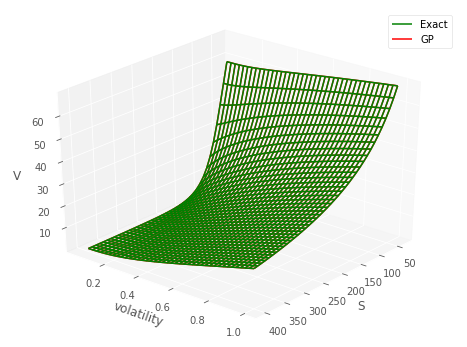}\\
(a) Error: $T-t=1.8$ & (b) Error: $T-t=1.0$ & (c) Error: $T-t=0.2$\\
\includegraphics[width=0.33\linewidth]{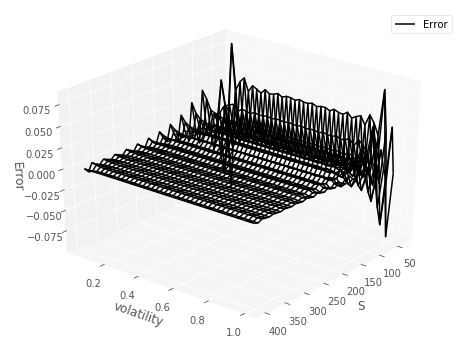} &
\includegraphics[width=0.33\linewidth]{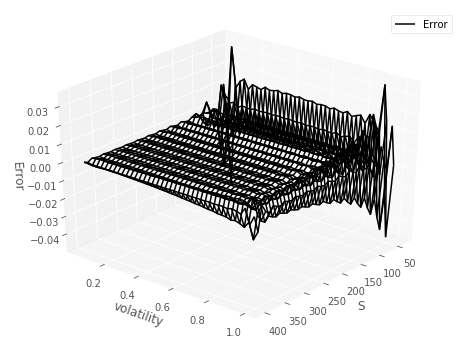} &
\includegraphics[width=0.33\linewidth]{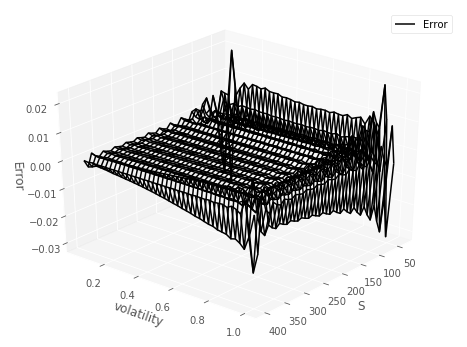}\\

\end{tabular}
\caption{\textit{Similar as Figure \protect\ref{fig:GPS_surface_call} for the put (instead of call) options.}}
\label{fig:GPS_surface_put}
\end{figure}

\begin{scriptsize}
\begin{lstlisting}[language=Python, caption=\textit{This Python 3.0 code excerpt illustrates how to use a GP to fit to option prices under a Heston model. $x_1$ and $x_2$ are gridded underlying stock values and volatilities respectively. Note that the listing provides the salient details only and the reader should refer to Example-6-GP-Heston.ipynb in Github for the full implementation.}, label=list:Heston] 
import PyHeston

S= 100
v0 = 0.1

lmbda = 0.1
meanV = 0.15
sigma = 0.1
r = 0.01
K = 100
T = 2.0
rho =-0.9
step_size = 0.4 #used internally by Heston pricer


lb = 1
ub = 400
portfolio = {}
portfolio['call']={}
portfolio['put']={}

training_number = 30
testing_number = 40  


x1_train = np.array(np.linspace(0.0,1.0, training_number), dtype='float32').reshape(training_number, 1)
x2_train = np.array(np.linspace(0.05,1.0, training_number), dtype='float32').reshape(training_number, 1)


X1_train, X2_train = np.meshgrid(x1_train, x2_train)
x_train = np.zeros(len(X1_train.flatten())*2).reshape(len(X2_train.flatten()), 2)
x_train[:,0] = X1_train.flatten()
x_train[:,1] = X2_train.flatten()

x1_test = np.array(np.linspace(0.0,1.0, testing_number), dtype='float32').reshape(testing_number, 1)
x2_test = np.array(np.linspace(0.05,1.0, testing_number), dtype='float32').reshape(testing_number, 1)

X1_test, X2_test = np.meshgrid(x1_test, x2_test)

x_test = np.zeros(len(X1_test.flatten())*2).reshape(len(X2_test.flatten()), 2)
x_test[:,0] = X1_test.flatten()
x_test[:,1] = X2_test.flatten()

portfolio['call']['price']= lambda x,y,z: PyHeston.HestonCall(lb+(ub-lb)*x, y, K, z, r, lmbda, meanV, sigma, rho, step_size)
portfolio['put']['price']= lambda x,y,z: PyHeston.HestonPut(lb+(ub-lb)*x, y, K, z, r, lmbda, meanV, sigma, rho, step_sizez)

for key in portfolio.keys():
    portfolio[key]['GPs'] = trainGPs(x_train, portfolio[key]['price'], timegrid)
    portfolio[key]['y_tests'], portfolio[key]['preds'], portfolio[key]['sigmas'] = predictGPs(x_test, portfolio[key]['price'], portfolio[key]['GPs'], timegrid)

\end{lstlisting}
\end{scriptsize}

\paragraph{Extrapolation} 

\begin{figure}[h!]
\centering
\begin{tabular}{cc}
\includegraphics[width=0.45\linewidth]{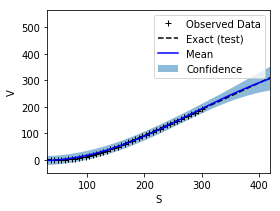} &
\includegraphics[width=0.45\linewidth]{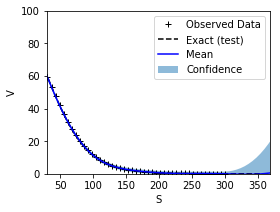}\\
(a) call price & (b) put price\\
\end{tabular}
\caption{\textit{This figure assesses the GP option price prediction in the  setup of a Black---Scholes model. The GP with a Linear and SE kernel is trained on $n=50$ $X,Y$ pairs, where $X\in \Omega^h\subset (0, 300]$ is the gridded underlying of the option prices and Y is a vector of call or put prices. These training points are shown by the black '+' symbols. The exact result using the Black-Scholes pricing formula is given by the black line. The predicted mean (blue solid line) and variance of the posterior are estimated from Equation \eqref{eq:posterior_moments} over $m=100$ gridded test points, $X_*\in \Omega^h_* \subset [300, 400]$, for the (left) call option struck at 110 and (center) put option struck at 90. The shaded envelope represents the 95\% uncertainty band about the mean of the posterior. This uncertainty band is observed to increase the further the test point is from the training set. The time to maturity of the options are fixed to two years.}}
\label{fig:extrapolation}
\end{figure}

One instance where kernel combination is useful in derivative modeling is for extrapolation---the appropriate mixture or combination of kernels can be chosen so that the GP is able to predict outside the domain of the training set. Noting that the payoff is linear when a call or put option is respectively deeply in and out-of-the money, we can configure a GP as a combination of a linear kernel and, say, a SE kernel. The linear kernel is included to ensure that prediction outside the domain preserves the linear property, whereas the SE kernel captures non-linearity. Figure \ref{fig:extrapolation} shows the results of using this combination of kernels to extrapolate the prices of a call struck at 110 and a put struck at 90. The linear property of the payoff function is preserved by the GP prediction and the uncertainty increases as the test point is further from the training set.

The above examples are trained on (semi)-analytic Black-Scholes and Heston prices, so the quality of the approximator can be assessed.

In a realistic application where approximators are trained on more general products in more general models, 
more complex, possibly approximate pricing schemes (including Monte Carlo inner simulations)
could be required
to find the values on the knot points.

\subsection{Greeking}

The GP provides analytic derivatives with respect to the input variables
\be
\partial_{X_*}\mean[f_* | X, Y, X_*] = \partial_{X_*}\muv_{X_*} + (\partial_{X_*} K_{X_*,X} )\alpha 
\label{eq:deriv}
\ee
where $\partial_{X_*} K_{X_*,X}= \frac{1}{\ell^2}(X-X_*)K_{X_*,X}$ and we recall from after \eqref{e:aloglik} that $\alpha=[K_{X,X} + \sigma^2 I]^{-1}Y$ (and in the numerical experiments we set $\muv=0$). Second order sensitivities are obtained by differentiating once more with respect to  $X_*$.

Note that $\alpha$ is already calculated at (pricing) training time  by Cholesky matrix factorization of $[K_{X,X} + \sigma^2 I]$ with $\mathcal{O}(n^3)$ complexity, so there is no significant computational overhead from Greeking. Once the GP has learned the derivative prices, Equation \eqref{eq:deriv} is used to evaluate the first order MtM Greeks with respect to the input variables over the test set. Example source code illustrating the implementation of this calculation is given in Listing \ref{list:deriv}.

\begin{figure}[h!]
\centering
\begin{tabular}{cc}
\includegraphics[width=0.45\linewidth]{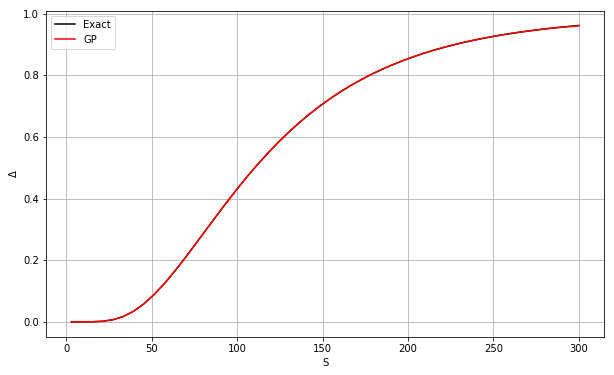} 
& 
\includegraphics[width=0.45\linewidth]{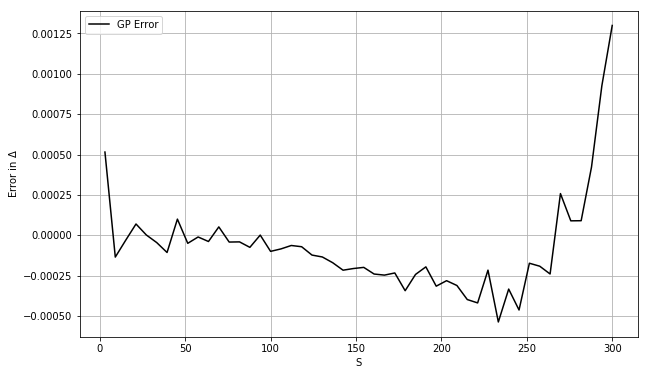} \\
\end{tabular}
\caption{\textit{This figure shows (left) the comparison of the GP estimate of the call option's delta $\Delta:=\frac{\partial C}{\partial S}$ and the BS delta formula. (Right) The error between the BS delta and the GP estimate.}}
\label{fig:delta}
\end{figure}

\begin{figure}[h!]
\centering
\begin{tabular}{cc}
\includegraphics[width=0.45\linewidth]{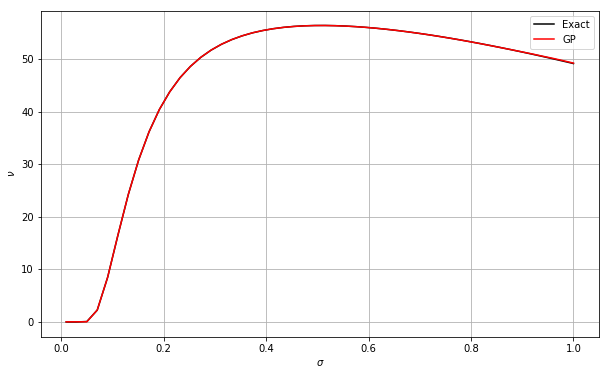} 
& 
\includegraphics[width=0.45\linewidth]{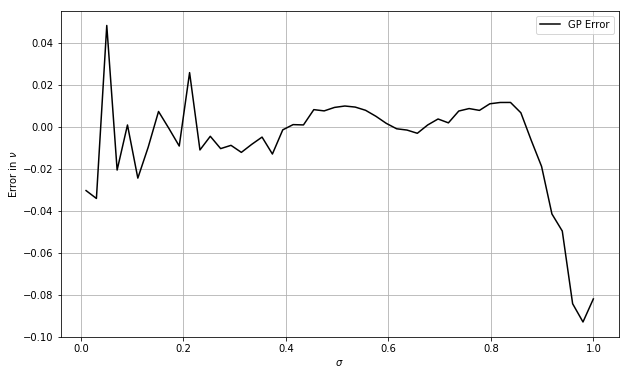} \\
\end{tabular}
\caption{\textit{This figure shows (left) the comparison of the GP estimate of the call option's vega $\nu:=\frac{\partial C}{\partial \sigma}$ and the BS vega formula. (right) The error between the BS vega and the GP estimate.}}
\label{fig:vega}
\end{figure}

Figure \ref{fig:delta} shows (left) the GP estimate of a call option's delta $\Delta:=\frac{\partial C}{\partial S}$ and (right) the error between the Black--Scholes (BS) delta and the GP estimate. We emphasize that the GP model is trained on underlying and option pricing data and not using the option's delta. The GP delta is observed to closely track the BS formula for the delta.
Figure \ref{fig:vega} shows (left) the GP estimate of a call option's
vega $\nu:=\frac{\partial C}{\partial \sigma}$, having trained on the implied volatility, and BS option model prices and not using the option's vega. The right hand pane shows the error between the BS vega and the GP estimate. The GP vega is observed to closely track the BS formula for the vega.

\begin{scriptsize}
\begin{lstlisting}[language=Python, caption=\textit{This Python 3.0 code excerpt, using scikit-learn, illustrates how to calculate the Greeks of an option by differentiating the GP price model. $x$ are gridded underlying stock values, so that $f\_prime$ is the estimate of the delta. If $x$ were gridded volatilities, then $f\_prime$ would be the estimate of the vega. The listing provides the salient details only and the reader should refer to Example-2-GP-BS-Derivatives.ipynb in Github for the full implementation.}, label=list:deriv] 
import scipy as sp
import numpy as np
from BlackScholes import *
from sklearn import gaussian_process
from sklearn.gaussian_process.kernels import ConstantKernel, RBF


# set BS model parameters
r = 0.0002  # risk-free rate
S= 100      # Underlying spot
KC = 130    # Call strike
KP = 70     # Put strike
sigma = 0.4 # implied volatility
T = 2.0     # Time to maturity
lb = 0.001  # lower bound on domain
ub = 300    # upper bound on domain
sigma_n = 1e-8 # additive noise in GP

call = lambda x,y: bsformula(1, lb+(ub-lb)*x, KC, r, T, y, 0)[0]
put = lambda x,y: bsformula(-1, lb+(ub-lb)*x, KP, r, T, y, 0)[0]

training_number = 100
testing_number = 50

x_train = np.array(np.linspace(0.01,1.2, training_number), dtype='float32').reshape(training_number, 1)
x_test = np.array(np.linspace(0.01,1.0, testing_number), dtype='float32').reshape(testing_number, 1)

y_train = []
    
for idx in range(len(x_train)):
    y_train.append(call(x_train[idx], sigma))
y_train = np.array(y_train)

sk_kernel = RBF(length_scale=1.0, length_scale_bounds=(0.01, 10000.0)) 
gp = gaussian_process.GaussianProcessRegressor(kernel=sk_kernel, n_restarts_optimizer=20)
gp.fit(x_train,y_train)
y_pred, sigma_hat = gp.predict(x_test, return_std=True)
    
l = gp.kernel_.length_scale
rbf= gaussian_process.kernels.RBF(length_scale=l)
 
Kernel= rbf(x_train, x_train)
K_y = Kernel + np.eye(training_number) * sigma_n
L = sp.linalg.cho_factor(K_y)
alpha_p = sp.linalg.cho_solve(np.transpose(L), y_train)
    
k_s = rbf(x_test, x_train)

k_s_prime = np.zeros([len(x_test), len(x_train)])
for i in range(len(x_test)):
    for j in range(len(x_train)):
        k_s_prime[i,j]=(1.0/l**2)*(x_train[j]-x_test[i])*k_s[i,j]
# Calculate the gradient of the mean using Equation in Greeking sub-section of paper.       
f_prime = np.dot(k_s_prime, alpha_p)/(ub-lb)

# show error between BS delta and GP delta
delta = lambda x,y: bsformula(1, lb+(ub-lb)*x, KC, r, T, y, 0)[1]
delta(x_test, sigma)-f_prime
\end{lstlisting}
\end{scriptsize}



\subsection{Mesh-Free GPs} \label{sect:mesh-free}

The above numerical examples have trained and tested GPs on uniform grids. This approach suffers from a stringent curse of dimensionality issue, as the number of training points grows exponentially with the dimensionality of the data (cf. Section \ref{ss:comp}). Hence, in practice, in order to estimate the MtM cube, we advocate divide-and-conquer, i.e.~the use of numerous low input dimensional space, $p$,  
GPs run in parallel on specific asset classes (see Section \ref{ss:scalcube} and \citep{GuhaniyogiLiSavitskySrivastava17}). 

Moreover, use of fixed grids is by no means necessary. We show here how GPs can show favorable approximation properties with a relatively small number of simulated reference points (cf. \citep{GramacyAple15}).

\begin{figure}[h!]
\centering
\begin{tabular}{cc}
\includegraphics[width=0.45\linewidth]{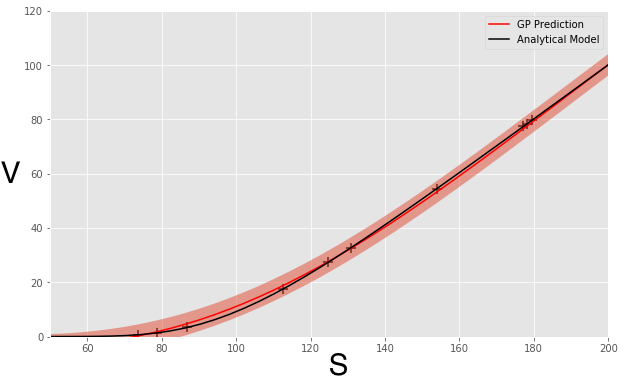}
& 
\includegraphics[width=0.45\linewidth]{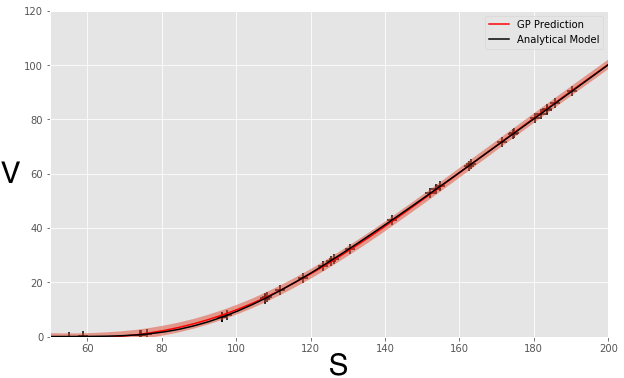} 
\end{tabular}
\caption{\textit{Predicted Heston Call prices using (left) 50 and (right) 100 simulated training points, indicated by '+'s, drawn from a uniform random distribution.}}  
\label{fig:meshfree}
\end{figure}

Figure \ref{fig:meshfree} shows predicted Heston call prices using (left) 50 and (right) 100 simulated training points, indicated by ``+''s, drawn from a uniform random distribution. The Heston call option is struck at $K=100$ with a maturity of $T=2$ years.
\begin{figure}[h!]
\centering
\begin{tabular}{cc}
\includegraphics[width=0.45\linewidth]{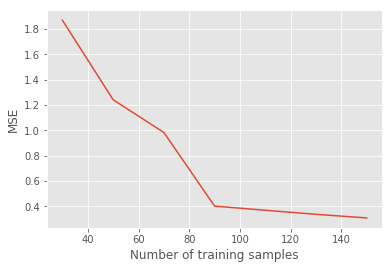} 
& 
\includegraphics[width=0.45\linewidth]{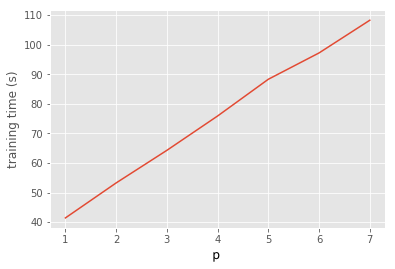} 
\end{tabular}
\caption{\textit{(Left) The convergence of the GP MSE of the prediction is shown based on the number of simulated Heston training points. (Right) Fixing the number of simulated points to 100, but increasing the dimensionality $p$ of each observation point (including more and more Heston parameters), the figure shows the wall-clock time for training a GP with SKI.}}  
\label{fig:scaling}
\end{figure}
Figure \ref{fig:scaling} (left) shows the convergence of the GP MSE of the prediction, based on the number of Heston simulated training points.

\subsection{Massively Scalable GPs} \label{sect:msgps}

 Fixing the number of simulated points to 100, but increasing the input space dimensionality, $p$, of each observation point (i.e. including more and more Heston parameters), Figure \ref{fig:scaling} (right) shows the wall-clock time for training a GP with SKI
(see Section \ref{ss:comp}). 
Note that the number of SGD iterations has been fixed to 1000.

\begin{figure}[h!]
\centering
\begin{tabular}{cc}
\includegraphics[width=0.45\linewidth]{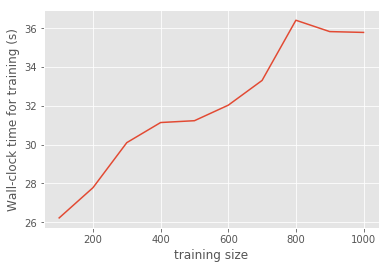} 
& 
\includegraphics[width=0.45\linewidth]{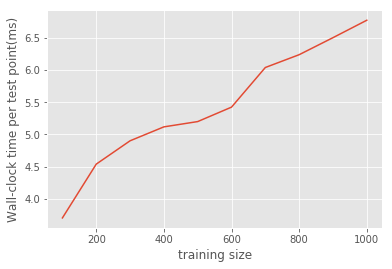} 
\end{tabular}
\caption{\textit{(Left) The elapsed wall-clock time is shown for training against the number of training points generated by a Black-Scholes model. (Right) The elapsed wall-clock time for prediction of a single point is shown against the number of testing points. The reason that the prediction time increases (whereas the theory reviewed in Section \ref{ss:comp} says it should be constant) is due to memory latency in our implementation---each point prediction involves loading a new test point into memory.}}  
\label{fig:MSGP}
\end{figure}

Figure \ref{fig:MSGP} shows the increase of MSGP training time and prediction time against the number of training points $n$ from a Black Scholes model. Fixing the number of inducing points to $m=30$
(see Section \ref{ss:comp}), 
we increase the number of observations, $n$, in the $p=1$ dimensional training set. 

Setting the number of SGD iterations to 1000, we observe an approximate 1.4 increase in training time for a 10x increase in the training sample. We observe an approximate 2x increase in prediction time for a 10x increase in the training sample. The reason that the prediction time grows with $n$ (instead of being constant, cf. Section \ref{ss:comp}) is due to memory latency in our implementation---each point prediction involves loading a new test point into memory. Fast caching approaches can be used to reduce this memory latency, but are beyond the scope of this research.

Note that training and testing times could be improved with CUDA on a GPU, but are not evaluated here.

\section{Multi-response Gaussian Processes} \label{sect:multiGP}
A multi-response Gaussian process is a collection of random vectors, any finite number of which have matrix-variate Gaussian distribution. We borrow from \citep{2017arXiv170304455C}
the following formulation
of a separable noise-free multi-response kernel specification as per \citep[Eq.~(9)]{AlvarezRosascoLawrence12}:

By definition, $\fv$ is a $d$ variate Gaussian process on $\mathbb{R}^p$ with vector-valued mean function $\bm{\mu} : \mathbb{R}^p\mapsto \mathbb{R}^d$, kernel $k: \mathbb{R}^p\times \mathbb{R}^p\mapsto \mathbb{R}$, and positive semi-definite parameter covariance matrix $\Omega \in \mathbb{R}^{d \times d}$, if the vectorization of any finite collection of vectors $\fv(\xv_1),\ldots,\fv(\xv_n)$ have a joint multi-variate Gaussian distribution,
 $$
\mathrm{vec}([\fv(\xv_1),\ldots,\fv(\xv_n)]) \sim \mathcal{N}(\mathrm{vec}(M), \Sigma \otimes \Omega),
 $$
 where $\fv(\xv_i) \in \mathbb{R}^d$ is a column vector whose components are the functions $\{\fv_l(\xv_i)\}_{l=1}^d$, $M$ is a matrix in  $\mathbb{R}^{d \times n}$ with $M_{li} = \mu_{l}(\xv_i)$,
$\Sigma$
is a matrix in  $\mathbb{R}^{n \times n}$
with $\Sigma_{ij} = k(\xv_i,\xv_j)$,
and $\Sigma \otimes \Omega$ is the Kronecker product 
$${\displaystyle  {\begin{pmatrix}\Sigma_{11}\Omega&\cdots &\Sigma_{1n}\Omega\\\vdots &\ddots &\vdots \\\Sigma_{m1}\Omega&\cdots &\Sigma_{mn}\Omega\end{pmatrix}}}.$$

Sometimes $\Sigma$ is called the column covariance matrix while $\Omega$ is the row (or task) covariance matrix. We denote $\fv \sim \mathcal{MGP}(\bm{\mu}, k, \Omega)$.
As explained after Eq.~(10) in \citep{AlvarezRosascoLawrence12}, the matrices $\Sigma$ and $\Omega$ encode dependencies among the inputs, respectively outputs.

\subsection{Multi-Output Gaussian Process Regression and Prediction with Noisy Observations}
In practice, the observations are not drawn from a function but exhibit noise. 
Given $n$ pairs of noisy observations $\{(\xv_i,\yv_i)\}_{i=1}^n, \xv_i \in \mathbb{R}^p, \yv_i \in \mathbb{R}^{d}$, we assume the model $\yv_i =\fv(\xv_i) + \mathbf{\epsilon},~i \in\{1,\ldots,n\}$, where
$\yv   \sim   \mathcal{MGP}(\bm{\mu},k',\Omega)$
with $k' = k(\xv_i,\xv_j) + \delta_{ij}\sigma^2$, in which $\sigma^2$ is the variance of an additive Gaussian i.i.d. noise, $\mathbf{\epsilon}$. 
The vectorization of the collection of functions $[\fv(\xv_1),\ldots,\fv(\xv_n)]$ therefore follows a multivariate Gaussian distribution
$$
\mathrm{vec}([\fv(\xv_1),\ldots,\fv(\xv_n)]) \sim \mathcal{N}(\bm{0},K_{X,X}\otimes\Omega),
$$
where $K_{X,X}$ is the $n \times n$ covariance matrix of which the $(i,j)$-th element $[K_{X,X}]_{ij} = k(\xv_i,\xv_j)$. 


To predict a new variable $\fv_* = [\fv_{*1},\ldots,\fv_{*m}]$ at the test locations $X_* = [\xv_{n+1},\ldots,\xv_{n+m}]$, the joint distribution of the training observations $Y = [\yv_1,\ldots,\yv_n]$ and the predictive targets $\fv_*$ are given by
\begin{equation}\label{joint}
  \begin{bmatrix}
  Y  \\
  \fv_*
  \end{bmatrix} \sim
  \mathcal{MN}    \left(
  \bm{0},
  \begin{bmatrix}
  K_{X,X}'   \quad K_{X_*,X}^T \\
  K_{X_*,X} \ \ K_{X_*,X_*}
  \end{bmatrix},
  \Omega   \right),
\end{equation}
where $K'_{X,X}$ is an $n \times n$ matrix of which the $(i,j)$-th element
$[K'_{X,X}]_{ij} = k'(x_{i},x_j)$, $K_{X_*,X}$ is an $m \times n$ matrix of which the $(i,j)$-th element
$[K_{X_*,X}]_{ij} = k(x_{n+i},x_j)$, and $K_{X_*,X_*}$ is an $m \times m$ matrix with
 the $(i,j)$-th element $[K_{X_*,X_*}]_{ij} = k(x_{n+i},x_{n+j})$. Thus, taking advantage of the conditional distribution of the multivariate Gaussian process, the predictive distribution is:
 \begin{equation}
   p(\mathrm{vec}(\bm{f}_*)|X,Y,X_*) =   \mathcal{N}(\mathrm{vec}(\hat{M}),\hat{\Sigma}\otimes\hat{\Omega}),
 \end{equation}
 where 
\begin{eqnarray}
  \hat{M}  &= &K_{X_*,X}^{T}(K'_{X,X})^{-1}Y, \\
  \hat{\Sigma}  &=& K_{X_*,X_*} - K_{X_*,X}^{\mathrm{T}}(K'_{X,X})^{-1}K_{X_*,X},\\
  \hat{\Omega} &= & \Omega  .
\end{eqnarray}

The hyperparameters and elements of the covariance matrix $\Omega$ are found by minimizing 
over $(\lambda, \Omega)$
the negative log marginal likelihood of observations: 
\begin{equation}\label{matrixLikelihood}
  \mathcal{L}(Y | X, \lambda, \Omega) = \frac{nd}{2}\ln(2\pi) + \frac{d}{2}\ln |K'_{X,X}| + \frac{n}{2}\ln |\Omega| + \frac{1}{2}tr((K'_{X,X})^{-1}Y\Omega^{-1}Y^{T}).
\end{equation}

Further details of the multi-GP are given in 
\citep{Bonilla2007}, \citep{AlvarezRosascoLawrence12}, and \citep{2017arXiv170304455C}. The computational remarks made in Section \ref{ss:comp} also apply here, with the additional comment that the training and prediction time also scale linearly (proportionally) with the number of dimensions $d$. Note that 
the task covariance matrix $\Omega$ is estimated via a $d-$vector factor $\mathbf{b}$ by $\Omega=\mathbf{b}\mathbf{b}^T +    \omega^2 I$ (where the
$\omega^2$ component corresponds to a standard white noise term).
An alternative computational approach, which exploits separability of the kernel, is described in Section 6.1 of \citep{AlvarezRosascoLawrence12}, with complexity $\mathcal{O}(d^3+n^3)$.

\subsection{Portfolio Value and Market Risk Estimation} \label{sect:portfolio}

The value $\pi$
of a portfolio of $d$ financial derivatives can 
typically 
be expressed as a linear combination of the components of a 
d-vector $\fv$
of a set of underlying risk factors $\xv$, i.e.
\begin{equation}\label{e:pi}
\pi(\xv)=\bm{w}^T\fv(\xv).
\end{equation}
We estimate the moments of the predictive distribution $p(\pi_* | X,Y,X_*)$ by
\begin{eqnarray}
  \mean[\pi_* | X, Y, X_*] &=& \bm{w}^T\hat{M} \label{pred_mean_pi}\\
  \mathrm{cov}(\pi_* | X, Y, X_*) &=& \bm{w}^{T}\hat{\Sigma}\otimes \Omega
  \bm{w}, 
  \label{pred_var_pi}
\end{eqnarray}
 where
\begin{eqnarray}
  \hat{M}  &= &K_{X_*,X}^{T}(K'_{X,X})^{-1}Y, \\
  \hat{\Sigma}  &=& K_{X_*,X_*} - K_{X_*,X}^{\mathrm{T}}(K'_{X,X})^{-1}K_{X_*,X}.
\end{eqnarray}
In particular, \eqref{pred_var_pi}
yields an expression for estimating the GP uncertainty in the point estimate of a portfolio, given the underlying risk factors, which accounts for the dependence between the financial derivative contracts. In general financial derivative contracts in the portfolio share common risk factors and the risk factors are correlated. Hence, a multi-GP approach, if not too demanding computationally, should be the meta-modelling method of choice for the $\{\fv_l(\xv_i)\}_{l=1}^d$.

Once the vector-function $\fv$ in \eqref{e:pi} has been learned, evaluating any portfolio spanned by $\fv$ becomes very fast. Hence the practical utility of a multi-GP approach is the ability to quickly predict 
portfolio values, together with an error estimate which also accounts for covariance of the derivative prices over the test points (conditional on the training points). 

Note that the meta-model only refers to $\fv$ (as opposed to the portfolio weights). Thus the predictive distribution of the portfolio remains valid even when the portfolio composition changes (e.g. in the context of
trade incremental XVA computations, see \citep[Section 5]{AlbaneseCrepey16}).  Note that, if a new derivative is added to the portfolio, we need not necessarily retrain all the GPs---the mean posterior estimate of the original portfolio value remains valid. However, the kernels must be relearned to update the covariance estimate. By construction, a derivative can be subtracted 
from the portfolio by simply setting the weight to zero---no retraining is required.


\paragraph{GP divide-and-conquer strategies} 

We reiterate that the benefit of using GPs is primarily for fast real-time computation. 

Since different GPs involved in the MtM cube computation are independent (cf. Section \ref{sect:mesh-free}), they can be trained in parallel over a grid of compute nodes such as a GPU or many-core CPU. In the case single-GPs are used, 
typically the number of input variables per model is small, and hence the training set consists of relatively few observations. The training of multi-GPs is more challenging since it involves fitting several instruments in the portfolio.

In practice, we can identify the subset of derivatives sharing common risk factors and fit a multi-GP to each subset. The computational overhead of multi-GP is justified by more accurate uncertainty estimates.

Instead of fitting a GP component (correlated with the others or not) to each derivative in a sub-portfolio as suggested above, an alternative can be to fit one  single-GP per overall sub-portfolio value. 
However,
if the weights of the portfolio are changed, then the corresponding GP must be re-trained.
We mention these pros and cons so that the most suitable approach can be assessed for each risk application.

\subsection{Numerical Illustration}\label{ss:numill}

\begin{figure}[h!]
\centering
\begin{tabular}{ccc}
\includegraphics[width=0.31\linewidth]{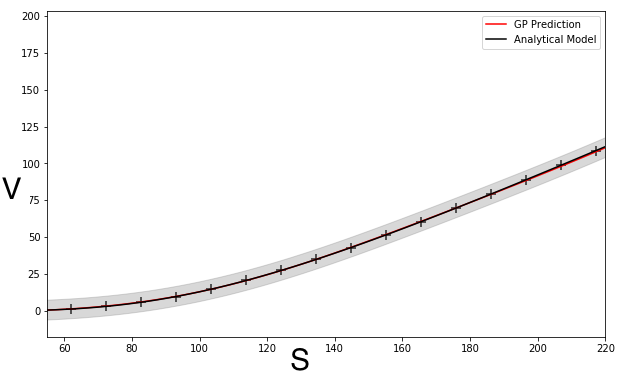} &
\includegraphics[width=0.31\linewidth]{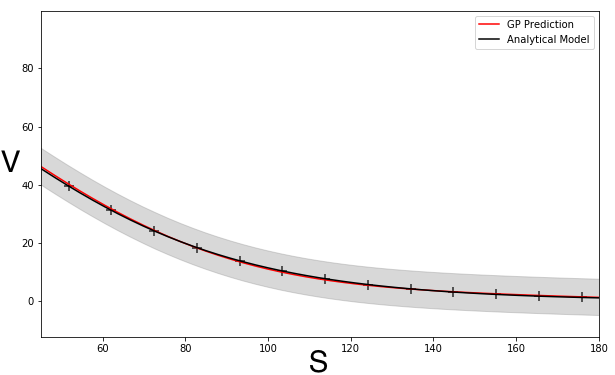} &
\includegraphics[width=0.31\linewidth]{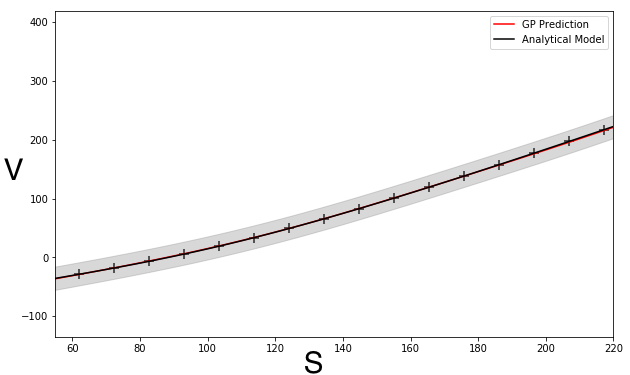}\\
(a) call price & (b) put price & (c) portfolio price\\
\end{tabular}
\caption{\textit{This figure compares the multi-GP option price prediction with the Black---Scholes model. The multi-GP with a MA kernel is trained on $n=50$ $X,Y$ pairs, where $X$ is the underlying gridded price and Y is a $d=2$-vector of corresponding call and put prices. The predicted mean (red line) and variance of the posterior are estimated from Equations \eqref{pred_mean_pi} and \eqref{pred_var_pi} over $m=100$ gridded test points, $S_*$, for the (left) call option (center) put option and (right) portfolio. The gray shaded envelope represents the 95\% confidence interval about the mean of the posterior. The exact result, using the Black-Scholes pricing formula, is given by the black line. The time to maturity of the options are fixed to two years.}}
\label{fig:portfolio}
\end{figure}

\begin{figure}[h!]
\centering
\begin{tabular}{cc}
\includegraphics[width=0.45\linewidth]{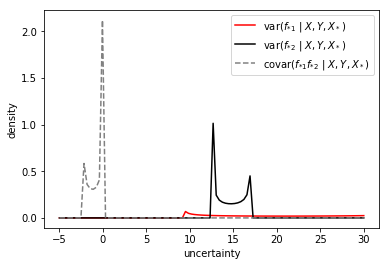} &
\includegraphics[width=0.45\linewidth]{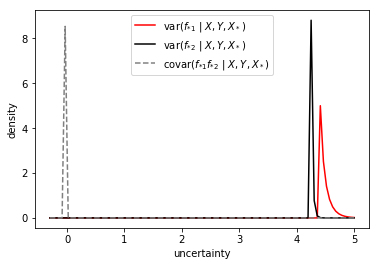} \\
(a) 5 training points & (b) 50 training points\\
\end{tabular}
\caption{\textit{This figure shows how multi-GPs are capable of attributing the uncertainty in the point estimate of the portfolio value to the constituent instruments. The multi-GP with a MA kernel is trained on $X,Y$ pairs, where $X$ is the gridded underlying of the option and Y is a $d=2$-vector of corresponding call and put prices. The elements of the posterior covariance matrix are plotted as a distribution over the test set of $m=100$ points. (left) Using 5 training points, we show the components of the posterior covariance. The cross terms in the posterior covariance, which are not given by single-GPs, are observed to be negative and material. (right) The posterior covariance with 50 training points. Each element is observed to shrink and is more homogeneous (i.e.~concentrated), leading to an overall reduction in portfolio value uncertainty, explaining the narrow homogeneous grey uncertainty band in Figure \ref{fig:portfolio}.}}
\label{fig:uncertainty}
\end{figure}

The above concepts are illustrated in Figures \ref{fig:portfolio} and \ref{fig:uncertainty} for 
a portfolio holding two long positions in a call option struck at 110 (left) and a short position in a put option struck at 90 (center), where $S_0=100$. Recall there is one risk factor which is common to both options---the underlying instrument $S$---and the maturity of each option is 2 years. 

To illustrate the uncertainty band under multi-GP regression, a bivariate-GP with a MA kernel (with $\nu$ fixed to 2.5) is trained to a Black-Scholes model as a function of $S$ on fifty training points, with additive Gaussian i.i.d. noise, as displayed in Listing \ref{list:portf}. Typically, one would use hundreds of training points. After 300 iterations the fitted kernel lengthscale and noise is $\hat{\lambda}=[0.208, 1.356355]$, and the fitted task covariance matrix is 
\begin{equation*}
\hat{\Omega}=\left[\begin{matrix}
36.977943 &  -1.1028435 \\
 -1.1028435 &  3.068603 
 \end{matrix}\right].
 \end{equation*}
 
The bivariate-GP subsequently estimates the values of the options and the portfolio at a number of test points. Some of these test points have been chosen to coincide with the training set and others are not in the set. The uncertainty in the point estimates is shown by the grey bands, denoting the 95\% GP uncertainty. In the portfolio case this uncertainty is a weighted combination of the uncertainty in the point estimate of each option price \emph{and} the cross-terms in the covariance matrix in Equation \eqref{pred_var_pi}. If, instead, single GPs were used separately for the put and the call price, then the uncertainty in the point estimate of the portfolio would neglect the cross-terms in the covariance matrix. 

To gain more insight into the components of the uncertainty, Figure \ref{fig:uncertainty} shows the distribution of uncertainty in the point estimates of
$\fv=(\fv_1,\fv_2)^{\sf T}$ over 100 testing points. Two experiments, with 5 and 50 training samples, are chosen to illustrate various properties in the multi-GP setting. The former experiment (left plot in figure) is chosen to highlight the importance of the cross-term in the posterior covariance $\textrm{cov}(\fv_{1*}, \fv_{2*}~|~X,Y,X_*)$, which is negative in this example. We reiterate that such a term is only represented in the multi-GP setting. In the case of noisy data or for a large portfolio,
it may yield a non-negligible contribution to the portfolio value uncertainty.

\begin{scriptsize}
\begin{lstlisting}[language=Python, caption=\textit{This Python 3.0 code excerpt, using GPyTorch, illustrates how to train a MC-GP for predicting the value of a toy portfolio containing a call and a put option (priced under Black-Scholes). We used the Adam update rule for SGD.
Note that the listing provides the salient details only and the reader should refer to Example-1-MGP-BS-Pricing.ipynb in Github for the full implementation.},label=list:portf]
import math
import torch
import gpytorch
import numpy as np
from scipy import *
from BlackScholes import *
from scipy import stats 

r = 0.0     # risk-free rate
S= 100      # Underlying spot
KC = 110    # Call strike
KP = 90     # Put strike
sigma = 0.3 # implied volatility
T = 2.0     # Time to maturity

call = lambda x: bsformula(1, lb+(ub-lb)*x, KC, r, T, sigma, 0)[0]
put = lambda x: bsformula(-1, lb+(ub-lb)*x, KP, r, T, sigma, 0)[0]

lb = 0.001           # lower bound on domain
ub = 300             # upper bound on domain
training_number = 50 # Number of training samples
testing_number = 100 # Number of testing samples

train_x = torch.linspace(0, 1.0, training_number)

train_y1 = torch.FloatTensor(call(np.array(train_x)))
train_y2 = torch.FloatTensor(put(np.array(train_x)))

#Create a train_y which interleaves the two
train_y = torch.stack([train_y1, train_y2], -1)

class MultitaskGPModel(gpytorch.models.ExactGP):
    def __init__(self, train_x, train_y, likelihood):
        super(MultitaskGPModel, self).__init__(train_x, train_y, likelihood)
        self.mean_module = gpytorch.means.MultitaskMean(
            gpytorch.means.ConstantMean(), num_tasks=2
        )
        
        self.covar_module = gpytorch.kernels.MultitaskKernel(gpytorch.kernels.ScaleKernel(
            gpytorch.kernels.MaternKernel(nu=2.5)), num_tasks=2, rank=1
        )
              
    def forward(self, x):
        mean_x = self.mean_module(x)
        covar_x = self.covar_module(x)
        return gpytorch.distributions.MultitaskMultivariateNormal(mean_x, covar_x)
        
test_x = torch.linspace(0, 1.0, testing_number)
test_y1 = torch.FloatTensor(call(np.array(test_x)))
test_y2 = torch.FloatTensor(put(np.array(test_x)))
test_y = torch.stack([test_y1, test_y2], -1)

likelihood = gpytorch.likelihoods.MultitaskGaussianLikelihood(num_tasks=2)
model = MultitaskGPModel(train_x, train_y, likelihood)

model.train()
likelihood.train()

# Use the adam optimizer
optimizer = torch.optim.Adam([
    {'params': model.parameters()},  # Includes GaussianLikelihood parameters
], lr=0.1)

# "Loss" for GPs---the marginal log likelihood
mll = gpytorch.mlls.ExactMarginalLogLikelihood(likelihood, model)

n_iter = 300
for i in range(n_iter):
    optimizer.zero_grad()
    output = model(train_x)
    loss = -mll(output, train_y)
    loss.backward()
    print('Iter %d/%d---Loss: %.3f lengthscale: %.3f' % (i + 1, n_iter, loss.item(), model.covar_module.data_covar_module.base_kernel.kernels[0].lengthscale))
    
    optimizer.step()

# Make predictions
model.eval()
likelihood.eval()
with torch.no_grad(), gpytorch.settings.fast_pred_var():
    y_hat= likelihood(model(test_x)) #Equation 15
lower, upper = y_hat.confidence_region() #Equation 16

# Fitted parameters
B = model.covar_module.task_covar_module.covar_factor.clone().detach()
v = model.covar_module.task_covar_module.var.clone().detach()
Omega = np.outer(B,B) + np.diag(v)
lengthscale=model.covar_module.data_covar_module.base_kernel.kernels[0].lengthscale
\end{lstlisting}
\end{scriptsize}

\section{CVA Computations} \label{sect:CVA}
In this section, as an example of a portfolio risk application, we consider the estimation of counterparty credit risk on a client portfolio. 
The expected loss to the bank associated with the counterparty defaulting is given by the (unilateral, see  \citep[Section 4.3]{AlbaneseCrepey16}) CVA.  
Taking discounted expectation of the losses triggered by the client default with respect to a pricing measure and the related discount process $\beta$, we obtain (assuming no collateral for simplicity)
\be \label{eq:CVA}
\textrm{CVA}_0=(1-R)\mathbb{E}\int_0^T  \beta_t 
\pi^+_{t}\delta_{\tau}(dt) ,
\ee
where $\delta_{\tau}$ is a Dirac measure at the client default time $\tau$ and $R$ is the client recovery rate. 

Assuming $\tau$ endowed with a stochastic intensity process $\gamma$ and a basic immersion setup between the market filtration and the filtration progressively enlarged with $\tau$ (see \citet[Section 8.1]{AlbaneseCaenazzoCrepey16a}), we have
\be \label{e:UCva}\textrm{CVA}_0 
=(1-R) \mathbb{E} \int_0^T  \beta_t \pi^+_{t} 
e^{-\int_0^t \gamma_s ds}  
 \gamma_t dt .\ee 

Under Markovian specifications, $\pi_t$ is a deterministic function of time and
 suitable risk factors $\bm{X}_t$, i.e.~$\pi_t=\pi(t,\bm{X}_t)$; likewise, in the case of intensity models,
 $\gamma_t=\gamma(t,\bm{X}_t)$. Factors common to $\pi$ and $\gamma$ allow modeling wrong way risk\footnote{Or, at least, soft wrong way risk, whereas hard wrong way risk may be rendered through common jump specifications (see \citep{CrepeySong15FS}).}, i.e.~the risk of adverse dependence between the risk of default of the client and the corresponding market exposure.

In the special case where the 
default is independent of the portfolio value expressed in numeraire units, then the expression \eqref{eq:CVA} simplifies to
\be \label{eq:CVAindep}
\textrm{CVA}_0=(1-R)\int_0^T\mathbb{E}[\beta_t\pi^+_{t} ]p(t) dt,
\ee
where  $p(t)$ is the probability density function of $\tau$. 
To compute the CVA numerically based on \eqref{eq:CVAindep} in this independent case, a set of $N$ dates 
$t_1,\ldots, t_N=T$ is chosen over which to evaluate the so-called expected positive exposure $\mathbb{E}[\beta_t \pi^+_t ]$. The probabilities $\Delta p_i=P(t_i\leq \tau < t_{i+1})$ can be bootstrapped from the CDS curve of the client (or some proxy if such curve is not directly available).

In stochastic default intensity models, one can evaluate likewise the $\mathbb{E}[ \beta_{t_i} \pi^+_{t_i} 
e^{-\int_0^{t_i}  \gamma_s ds}  
 \gamma_t ]$ and compute the CVA based on \eqref{e:UCva}, or simulate $\tau$ and compute the CVA based on \eqref{eq:CVA}.
 
Note that the portfolio weights $w_i$ in \eqref{e:pi} are all 0 or 1 in the context of trade incremental XVA computations (cf. \citep[Section 5]{AlbaneseCrepey16}).

 The above approximation uses Gaussian process regression to provide a fast approximation for
$\pi$ valuation. A metamodel for $\pi$ is fitted to model generated data, assuming a data generation process for the risk factors. Our GP regression provides an estimation of the GP error in the point estimate of the portfolio value (also accounting for the dependence between the 
portfolio ingredients, i.e.~between the $\fv_l$ in \eqref{e:pi}, provided multi-GP is used).

Hence, we use machine learning to learn the component derivative exposures as a function of the underlying and other parameters, including  (by slicing in time) time to maturity. The ensuing CVA computations are then done by Monte Carlo simulation based on this metamodel for $\pi$.
This procedure is referred to as MC-GP CVA computational approach hereafter.
It saves one level of nested (such as inner Monte-Carlo) full revaluation (referred to as MC-reval hereafter), while avoiding parametric regression schemes for $\pi$ (at each $t_i$), which have little adaptivity and error control.


\subsection{MC-GP Estimation of CVA}

First we consider the independent case  \eqref{eq:CVAindep},
which entails the following Monte-Carlo estimate of the CVA over M paths, along which the market risk factors are sampled:
\be
\textrm{CVA}_0 \approx \frac{(1-R){\Delta t}}{M}\sum_{j=1}^M \sum_{i=1}^{N}\pi(t_i,\bm{X}^{(j)}_{t_i})^+ \beta^{(j)}_{t_i}\Delta p_i,
\ee
where the exact portfolio value $\pi(t_i, \bm{X}^{(j)}_{t_i})^+$ is evaluated for simulated risk factor $\bm{X}^{(j)}_{t_i}$ in path $j$ at time $t_i$. 

Then we replace the exact portfolio value by the mean of the posterior function conditioned on the simulated market risk factors $\Xv_{t_i}$, which results in the following CVA estimate (assuming a uniform time-grid with step $\Delta t$):

\begin{eqnarray*}
\widehat{\textrm{CVA}}_0  &=& \frac{(1-R){\Delta t}}{M}\sum_{j=1}^M \sum_{i=1}^{N} \beta^{(j)}_{t_i}\big(\mean[\pi_*| X, Y, \xv_*= \Xv^{(j)}_{t_i}]\big)^+\Delta p_i
\label{eq:MC-GP} 
\end{eqnarray*}

 
In the stochastic intensity case \eqref{e:UCva},
the above formulas become 
\be
\textrm{CVA}_0 \approx \frac{(1-R) \Delta t }{M}\sum_{j=1}^M \sum_{i=1}^{N}\beta^{(j)}_{t_i}\pi(t_i,\bm{X}^{(j)}_{t_i})^+ e^{-\Delta t\sum_{\iota<i}\gamma(t_\iota,\bm{X}^{(j)}_{t_\iota}) }\gamma(t_i,\bm{X}^{(j)}_{t_i})
\ee
and
\beql{e:cva-sto}
 &  \widehat{\textrm{CVA}}_0 = \frac{(1-R) \Delta t }{M}\sum_{j=1}^M \sum_{i=1}^{N}
\beta^{(j)}_{t_i}\big(\mean[\pi_*| X, Y, \xv_*= \Xv^{(j)}_{t_i}]\big) ^+
\\& \qqq\qqq\qqq e^{-\Delta t\sum_{\iota<i}\gamma(t_\iota,\bm{X}^{(j)}_{t_\iota}) }\gamma(t_i,\bm{X}^{(j)}_{t_i}).
\eeql
The MC sampling error in the GP-MC estimate of the CVA is given by
\beql{eq:cva-err}
&
 \frac{1}{M-1}\sum_{j=1}^M \Big[(1-R) \Delta t \sum_{i=1}^{N}
\beta^{(j)}_{t_i}\big(\mean[\pi_*| X, Y, \xv_*= \Xv^{(j)}_{t_i}]\big) ^+
\\& \qqq \qqq \qqq e^{-\Delta t\sum_{\iota<i}\gamma(t_\iota,\bm{X}^{(j)}_{t_\iota}) }\gamma(t_i,\bm{X}^{(j)}_{t_i})- \widehat{\textrm{CVA}}_0\Big]^{2}.
\eeql

In the context of CVA on equity or commodity derivatives, structural default models may be found more suitable than default intensity models (see e.g. \citep{BallottaGianluca15}).
A Monte Carlo GP approach is then still workable, by relying on the native formulation \eqref{eq:CVA} of the CVA. The latter can be implemented based on client default simulation, which does not require that the client default time has an intensity.

\subsection{Expected Positive Exposure Profile and Time 0 CVA}\label{ss:epecva}
 
We continue with the 
same portfolio and option model (for data generation) as in the 
example of Section \ref{ss:numill} (of course,
 in practice XVAs are mainly for OTC derivatives, instead of  
 exchange tradeable options in this example, but our purpose is purely expository).  Table \ref{tab:sim_params_epe} shows the values for the Euler time stepper used for simulating Black-Scholes dynamics over a two year horizon.
\begin{table}[h!]
\centering
\begin{tabular}{|l|cc|}
\hline
Parameter description & Symbol & Value \\
\hline
Number of simulations & $M$ & 1000\\
Number of time steps & $N$ & 100\\
Initial stock price & $S_0$ & 100 \\
\hline
\end{tabular}
\caption{\textit{This table shows the values for the Euler time stepper used for market risk factor simulation.}}
\label{tab:sim_params_epe}
\end{table}
 
\begin{figure}[h!]
\centering
\begin{tabular}{cc}
\includegraphics[width=0.45\linewidth]{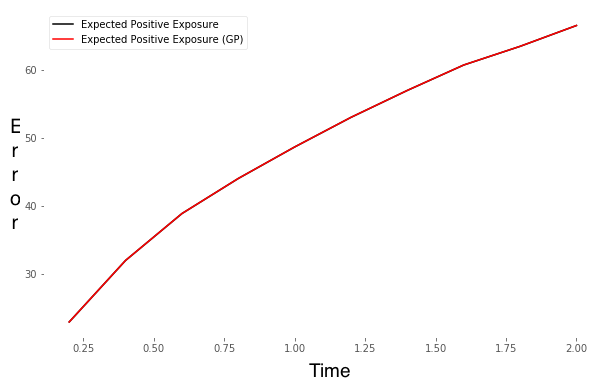} 
& 
\includegraphics[width=0.45\linewidth]{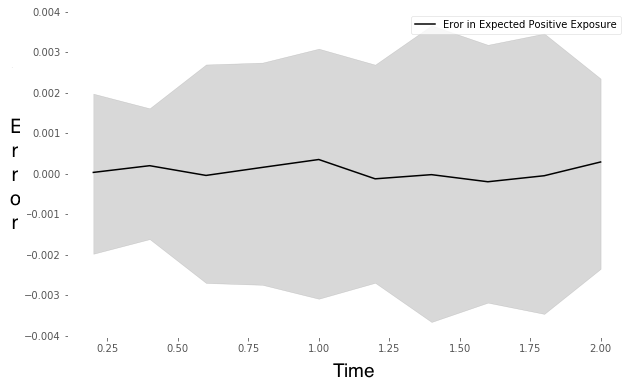} 
\end{tabular}
\caption{\textit{(Left) MC-reval and MC-GP estimates of the EPE of the portfolio over time (the two graphs are practically indistinguishable). (Right) The error in the MC-GP estimate of the EPE (black) and  (grey) the 95\% GP uncertainty band are also shown against time.}}
\label{fig:GPS_EPE}
\end{figure}

Figure \ref{fig:GPS_EPE} compares the (left) MC-reval (i.e., with full reevaluation of the portfolio, in this case by the Black Scholes formula) and MC-GP estimate of $\mathbb{E}(\pi^+_t)$, the expected positive exposure (EPE) of the portfolio, over time.
The error in the MC-GP estimate and 95\% uncertainty band, exclusive of the MC sampling error, is also shown against time (right).

In order to illustrate CVA estimation using both credit and market simulation, we introduce the following dynamic pre-default intensity (cf. \citep{10.1007/978-3-0348-0097-6_16}):
\be \label{eq:credit}
{\gamma}(S_t)={\gamma}_0(\frac{S_0}{S_t})^{{\gamma}_1},
\ee
where $({\gamma}_0, {\gamma}_1)=(0.02,1.2)$.
The time 0 CVA is then computed based on \eqref{e:cva-sto} as displayed in Listing \ref{list:CVA}.

\begin{figure}[h!]
\centering
\includegraphics[width=0.8\linewidth]{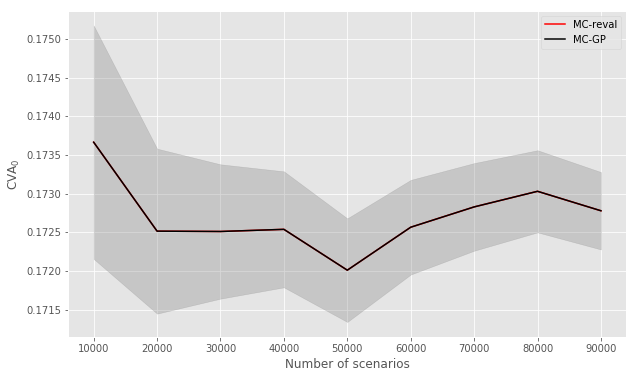} 
\caption{\textit{This figure shows the Monte Carlo convergence properties of the MC-GP CVA$_0$ estimate. Using 100 training samples, the MC-GP CVA$_0$ is estimated over an increasingly larger number of test samples. The 95\% Monte Carlo confidence interval in the 
GP-MC CVA$_0$ is also shown by the gray band centered around the MC-GP CVA$_0$ point estimate. Also shown is the MC-reval CVA$_0$ estimate, which is practically indistinguishable from the other---the difference in the GP CVA$_0$ estimate and the CVA$_0$ with revaluation is much smaller than the MC sampling error.}}
\label{fig:GPS_CVA}
\end{figure}

Setting $R=40\%$ hereafter,
Figure \ref{fig:GPS_CVA} shows how the standard error in the MC-GP CVA$_0$ estimate versus MC-reval decays against the number of training samples used for each GP model. The 95\%  uncertainty band of the GP prediction is also shown. 

\begin{scriptsize}
\begin{lstlisting}[language=Python, caption=\textit{This Python 3.0 code excerpt, using scikit-learn, illustrates how to simulate the time 0 CVA of a portfolio using MC-GP.   The implementation assumes BS pricing with a dynamic default intensity model given by Equation \eqref{eq:credit}. Note that the listing provides the salient details only and the reader should refer to Example-3-MC-GPA-BS-CVA.ipynb in Github for the full implementation.}, label=list:CVA]
def CVA_simulation(sim_params, model_params, def_model):
    
    n_sim_dt = sim_params['n_sim_dt'] # number of Euler stpes
    M        = sim_params['M']        # number of paths
    nt       = sim_params['nt']       # number of exposure dates
    timegrid = sim_params['timegrid'] # time grid of exposure dates
    r        = model_params['r']
    sigma    = model_params['sigma']
    T        = model_params['T'] 
    t0       = model_params['t0'] 
    S0       = model_params['S0'] 
    gamma_0  = def_model['gamma_0']
    gamma_1  = def_model['gamma_1']
    
    
    stride = n_sim_dt/(nt-1)
    idx = np.arange(0,n_sim_dt+1,stride, dtype=int)
    
    pi = {}
    pi['tilde'] = np.array([0.0]*(nt-1)*M, dtype='float32').reshape((nt-1), M)     # GP portfolio value
    pi['exact'] = np.array([0.0]*(nt-1)*M, dtype='float32').reshape((nt-1), M)     # BS portfolio value
    pi['tilde_var'] = np.array([0.0]*(nt-1)*M, dtype='float32').reshape((nt-1), M) # GP portfolio variance
    gamma = np.array([0.0]*(nt-1)*M, dtype='float32').reshape((nt-1), M)           # hazard rates
    dPD = np.array([0.0]*(nt-1)*M, dtype='float32').reshape((nt-1), M)             # default probabilities
    
    #simulate underlying Black-Scholes dynamics using Euler
    S = gbm(S0, r, sigma, T-t0, n_sim_dt, M)
    
    if (def_model['calibrate']):
        x  = np.exp(S0/S)**gamma_1
        # default probability (assumed to be estimated from credit spread)
        dt = timegrid[1]-timegrid[0] 
        f  = lambda y: np.abs(np.mean(np.prod(x**(-y*dt), axis=0))---def_model['p']) 
        res = sp.optimize.basinhopping(f, 0.1, niter=10)
        i = 1   
        while (abs(res.fun) >1e-3):   
          res = sp.optimize.basinhopping(f, 0.1, niter=100*i)
          i *= 2
        gamma_0= res.x[0]
        print("calibration:", gamma_0, gamma_1, f(gamma_0), res.fun)  
    
    
    for m in range(M):  
      i = 1 
      exp_factor=1
        
      for time in timegrid[1:]:
        dt = timegrid[i]-timegrid[i-1] 
        
        S_= S[idx[i],m] # simulated S
        # avoid simulated S breaching boundaries of domain
        if (S_<lb):
            mins=S_
            S_=lb
        if (S_>ub):
            S_=ub
            maxs=S_
    
        pred_= 0
        v_ = 0
        var_ =0 
    
        for key in portfolio.keys():
           pred, std = portfolio[key]['GPs'][i].predict(np.array([(S_-lb)/(ub-lb)]).reshape(1,-1),return_std=True) 
           pred_ += portfolio[key]['weight']*pred
           var_ += (portfolio[key]['weight']*std)**2 
       
           if key=='call':
              v_ += portfolio[key]['weight']*bsformula(1, S_, KC, r, time, sigma, 0)[0]
           else:
              v_ += portfolio[key]['weight']*bsformula(-1, S_, KP, r, time, sigma, 0)[0]
        pi['tilde'][i-1,m] = np.maximum(pred_,0)
        pi['exact'][i-1,m] = np.maximum(v_,0)
        pi['tilde_var'][i-1,m] =var_ 
          
        # default intensity model
        gamma[i-1,m] = gamma_0*(S0/S_)**gamma_1    
        
        # compute default probabilities  
        exp_factor*=np.exp(-dt*gamma[i-1,m])    
        dPD[i-1,m]= gamma[i-1,m]*exp_factor
        
        i += 1
    # compute CVA
    i = 0
    CVA ={}
    CVA['tilde'] = 0
    CVA['exact'] = 0
    CVA['tilde_up'] = 0
    CVA['tilde_down'] = 0
    CVA['var_tilde'] =0
   
    for time in timegrid[1:]:
        dt = timegrid[i+1]-timegrid[i]
        CVA['tilde'] += np.mean(dPD[i,:]*pi['tilde'][i,:])*np.exp(-r*(time-t0))*dt
        CVA['var_tilde'] += np.var(dPD[i,:]*pi['tilde'][i,:]*np.exp(-r*(time-t0))*dt)
        CVA['exact'] += np.mean(dPD[i,:]*pi['exact'][i,:])*np.exp(-r*(time-t0))*dt
        i+=1
    
    CVA['tilde_up'] = (1-def_model['recovery'])*(CVA['tilde'] + 2*np.sqrt(CVA['var_tilde']/M))
    CVA['tilde_down'] = (1-def_model['recovery'])*(CVA['tilde']---2*np.sqrt(CVA['var_tilde']/M))
    CVA['tilde'] *= (1-def_model['recovery'])
    CVA['exact'] *= (1-def_model['recovery'])
        
    return(CVA)
\end{lstlisting}
\end{scriptsize}

\subsection{Incremental One-Year CVA VaR}
In this section, we demonstrate the application of GPs to the estimation of the Value-at-risk (VaR, i.e.~quantile) of level $\alpha$ of the one year incremental CVA. The purpose of the calculation is to estimate, at the confidence level $\alpha$, the extent to which to CVA liability of a bank may increase over the next year. For this purpose, we estimate the distribution of the incremental CVA over one year, i.e.~of the random variable
($\textrm{CVA}_1 -\textrm{CVA}_0$).

For the purpose of illustration, we again use the dynamic pre-default intensity in \eqref{eq:credit},
 for fixed parameters ${\gamma}_0=0.02$ and ${\gamma}_1=1.2$ in \eqref{eq:credit},
 and the same market portfolio as before. However, we now model the (pre-default) CVA process such that (with  zero interest rates)
\be
\mathbf{1}_{t<\tau}{\rm CVA}(t,  S_t)=\mathbf{1}_{t<\tau}\mathbb{E}[\mathbf{1}_{\tau<T}\big(\pi(\tau, S_\tau )\big)^+~|~S_t , {t<\tau}].
\ee
We fix the pre-default intensity model parameters. Overall, the MC-GP estimation of  VaR$(\textrm{CVA}_1-\textrm{CVA}_0)$ is implemented as a nested simulation, with an outer loop over the simulation of the underlying out to one year, and a nested MC simulation for the point estimation of the one year CVA along each path. The CVA$_0$, by contrast, is estimated with only an outer simulation loop and is a non-negative scalar.   

\begin{figure}[h!]
\centering
\begin{tabular}{cc}
\includegraphics[width=0.45\linewidth]{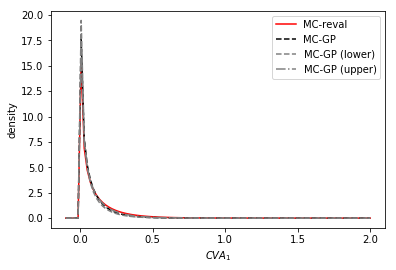} 
&
\includegraphics[width=0.45\linewidth]{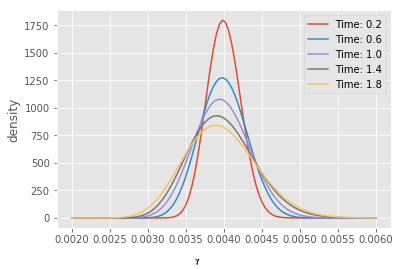} \\

\end{tabular}
\caption{\textit{(Left) The distribution of CVA$_1$, as estimated by MC-reval (MC with full repricing using Black-Scholes formulas) versus a MC-GP method with 95\% confidence intervals. The default model uses fixed parameters ${\gamma}_0=0.02$ and ${\gamma}_1=1.2$. In order to isolate the effect of the GP approximation, we use identical random numbers for each method.  (Right) The distribution of $\gamma\equiv{\gamma}(S_t)$ at various times over the simulation horizon for a fixed parameters: ${\gamma}_0=0.02$, ${\gamma}_1=1.2$.}}  
\label{fig:GPS_EPE-bis}
\end{figure}

Figure \ref{fig:GPS_EPE-bis} (left) shows the distribution of CVA$_1$, as estimated with a MC-reval (i.e.~using Black-Scholes formulas at time 1) or a MC-GP method. Also, not shown, CVA$_0=0.2$, and hence the random variable ($\textrm{CVA}_1 -\textrm{CVA}_0$) can be negative.  
In order to isolate the effect of the GP approximation, we use identical random numbers for each method. 

The MC-GP and MC-reval graphs are practically indistinguishable from each other. The reason for sharp approximation is three-fold: 
(i) the dimension (in the sense of the number of risk factors) is only 1, (ii) the statistical experiment has been configured as an interpolation problem, with many of the gridded training points close to the gridded test points; and (iii) the training sample size of 200 is relatively large to approximate smooth surfaces (with no outliers).  The right hand plot shows the distribution of $\gamma\equiv{\gamma}(S_t)$ at various times over the simulation horizon.

\subsection{CVA Uncertainty Quantification}

In this section, we demonstrate the application of GPs to the uncertainty quantification of CVA computations, given a prior on the credit risk model parameters.

Namely, the parameters, $({\gamma}_0, {\gamma}_1)$, of the dynamic intensity model \eqref{eq:credit} are now in one-to-one correspondence through the constraint 
\be \label{eq:constraint}
\mathbb{E} e^{-\int_0^T {\gamma}(S_t) dt}=\mathbb{P} (\tau>T),
\ee
in which the right-hand side is a given target value  extracted from the client CDS curve  (or a suitable proxy for the latter).  
Instead of fixing $(\gamma_0, \gamma_1)$, we now place a chi-squared prior over $\gamma_1$, which is centered at $1.2$. For each sample of $\gamma_1$,  the corresponding value of $\gamma_0$ is determined by time discretization and numerical root finding through \eqref{eq:constraint}). 

\begin{figure}[h!]
\centering
\begin{tabular}{ccc}
\includegraphics[width=0.33\linewidth]{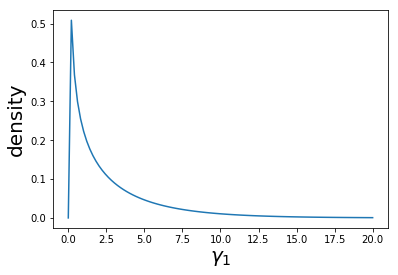} &
\includegraphics[width=0.33\linewidth]{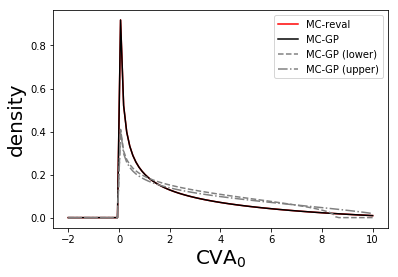} &
\includegraphics[width=0.33\linewidth]{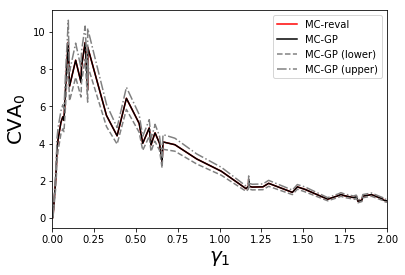} \\
\end{tabular}
\caption{\textit{The default model uses 1000 simulated parameter values with a chi-squared prior centered at ${\gamma}_1=1.2$ (left) and ${\gamma}_0$ is found by a constrained optimization.
(center) The density of the time 0 CVA posterior, as estimated by MC-reval (MC with full repricing using Black-Scholes formulas) versus a MC-GP method
(with 95\% Monte Carlo confidence intervals centered about the MC-GP estimates). 
(right) CVA$_0$ is shown against $\gamma_1$ with 95\% Monte Carlo confidence intervals centered about the MC-GP estimates. }}  
\label{fig:CVA_0_UQ}
\end{figure}

Figure \ref{fig:CVA_0_UQ} shows the density of the time 0 CVA posterior, as estimated by MC-reval (MC with full repricing using Black-Scholes formulas) versus 
MC-GP. 

Next we show the estimation of the one year CVA VaR with uncertainty quantification. As displayed in Listing \ref{CVAVaR}, the MC-GP estimation is implemented as a doubly nested simulation, with an outer loop for the sampling from the prior distribution on $\gamma_1$, a middle nested loop for simulation of the underlying out to one year, and an inner nested MC simulation for the point estimation of the one year CVA along each path.

\begin{figure}[h!]
\centering

\includegraphics[width=0.7\linewidth]{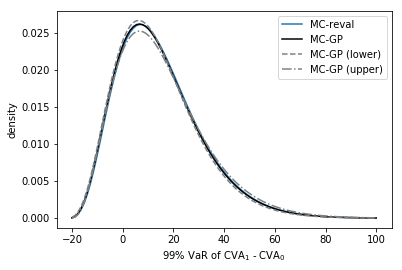} \\

\caption{\textit{This figure shows the distribution of the 99\% VaR of $({\rm CVA}_1-{\rm CVA}_0)$ 
under a chi-squared prior on ${\gamma}_1$ in Equation \eqref{eq:credit}  and prior on ${\gamma}_0$ which satisfies the constraint \eqref{eq:constraint} with $\mathbb{P} (\tau>2)=0.05$. 1000 outer-simulations are used for sampling from the prior on ${\gamma}_1$. The MC-GP and MC-reval 99\% CVA VaRs are observed to be practically identical under the same random numbers}.} 
\label{fig:GPS_VaR}
\end{figure}

Figure \ref{fig:GPS_VaR} shows the distribution of the 99\% VaR of (${\rm CVA}_1- {\rm CVA}_0$) under a chi-squared prior on the parameter ${\gamma}_1$ and the corresponding value of ${\gamma}_0$ is found from solving \eqref{eq:constraint} with $\mathbb{P} (\tau>2)=0.05$. The MC-GP and MC-reval 99\% CVA VaRs are observed to be practically identical under the same random numbers.

\begin{scriptsize}
\begin{lstlisting}[language=Python, caption=\textit{This Python 3.0 code excerpt, using scikit-learn, illustrates how to simulate the CVA-VaR of a portfolio using MC-GP. The implementation assumes a Black--Scholes model with a dynamic default intensity model given by Equation \eqref{eq:credit}. The BS parameters and portfolio configuration are the same as the previous listing. Note, for conciseness, that this excerpt should only be run after running the previous excerpt. Note that the listing provides the salient details only and the reader should refer to Example-4-MC-GPA-BS-CVA-VaR.ipynb in Github for a complete implementation.},label=CVAVaR]
J= 1000 # number of outer simulations (from prior)
M = 1000 # number of middle simulations
CVA = []
CVA_0 = []
gamma_1= np.array([0.0]*J, dtype='float32')

# Sample from prior distribution using 
# non-centered chi-squared random variates
gamma_1 = (1.2 + 1.0*np.random.randn(J))**2 

for j in range(J): # outer loop
    
    def_model['gamma_0'] = 0.02
    def_model['gamma_1'] = gamma_1[j] 
    model_params['t0'] = 0.0
    sim_params['timegrid'] = timegrid
    CVA_0.append(CVA_simulation(sim_params, model_params, def_model))
    
    S = gbm(S0, r, sigma, 1.0, n_sim_dt, M)
    model_params['t0'] = 1.0
    sim_params['timegrid'] = timegrid[5:]
    for m in range(M): # middle loop
        model_params['S0'] = S[-1, m]
        CVA.append(CVA_simulation(sim_params, model_params, def_model))
\end{lstlisting}
\end{scriptsize}
 
\subsection{Scalability of the Approach}\label{ss:scalcube}

We demonstrate the application of our GP meta-model to the CVA estimation on a counterparty portfolio of interest rate swaps (IRSs). The portfolio holds both short and long positions in 20 IRSs on a total of eleven interest rates and 10 FX rates. The contracts range from 5 to 10 years in maturity. 

The short interest and FX rates data are generated from mean reverting processes with a quasi-homogeneous
correlation structure between the driving Brownian motions of our factors. 
Specifically, we
set the correlation between interest rates to 0.45, between interest rates \& FX to 0.3 and 0.15 between FX rates. In our experiment, $n=10$, giving a total of 21 factors.

We use the multi-factor Hull-White model for the short rates. For completeness the details of the models are given in Appendix \ref{sect:rates}.

These models are simulated under a Euler scheme with a time step of 0.01. Every ten time steps, i.e. for a coarse $\Delta t=0.1$, we store the simulated rates and evaluate the IRS prices. 

The IRS contracts are spot-starting, with the first reset at time $t_0$ and subsequent resets every $\delta =0.5$ years at times $\left\{t_n\right\}_{0 \leq n \leq N}$. To price the swap,  let's denote by $L(t,T)$ the simple rate prevailing at time t for a maturity T, which is deterministic function of the short rate given by the Hull-White model.  Then the foreign price of the IRS at reset dates to the party receiving floating and paying fixed is given in units of notional denominated in the foreign currency as:
\be
p\left(t_n\right) = 1+\delta L\left(t_{n-1}, t_n\right) - \frac{1}{1+\left(N-n\right) \delta L\left(t_n, t_N\right)} - \delta S \sum_{i=0}^{N-n} \frac{1}{1+i \delta L\left(t_n, t_{n+i}\right)}, ~ \forall 1 \leq n \leq N,
\ee
and
\be
p\left(t_0\right) = 1 - \frac{1}{1+N \delta L\left(t_0, t_N\right)} - \delta S \sum_{i=1}^{N} \frac{1}{1+n \delta L\left(t, t_n\right)}.
\ee
For non-reset dates, i.e. any $t \in ]t_n, t_{n+1}[, 0\leq n \leq N-1$ we have:

\be 
p\left(t\right) = \frac{1+\delta L\left(t_n, t_{n+1}\right)}{ 1+\left(t_{n+1}-t\right) L\left(t, t_{n+1}\right) }- \frac{1}{1+ \left(t_N-t\right) \delta L\left(t, t_N\right)} - \delta S \sum_{i=1}^{N-n} \frac{1}{1+\left(t_{n+i}-t\right) L\left(t, t_{n+i}\right)}.
\ee
Except for $t=t_0$, we note that the price $p\left(t\right)$ is always a function of both the short rate at $t$ and at the previous reset date.

Each GP model therefore uses up to three inputs to learn the swap price: the domestic or foreign short rate, the same rate at the latest reset date, and if foreign the FX rate into Euros. Hence a GP model for a domestic interest rate swap price has two input variables and a foreign interest rate has three.

To train each GP model, we use 1000 paths of short and FX rates and Euro denominated mark-to-market IRS prices at the coarse time step $\Delta t=0.1$. Each GP model is trained as a surrogate of each IRS contract in the portfolio. With ten periods, there are therefore 200 GP models trained. 

The above numerical examples have trained and tested GPs on uniform grids. This approach suffers from a stringent curse of dimensionality issue, as the number of training points grows exponentially with the dimensionality of the data (cf. Section \ref{ss:comp}). Hence, in practice, in order to estimate the MtM cube, we advocate divide-and-conquer, i.e.~the use of numerous low input dimensional space, $p$,   
GPs run in parallel on specific asset classes (see Section \ref{sect:portfolio}). In the present example, we even
train one GP per instrument.  
As seen in the end of Section \ref{sect:portfolio},
the advantage of this extreme case of a divide-and-conquer approach is that the portfolio can then be rebalanced without the need to retrain the GP. 

All variables, including the prices, are rescaled to the unit interval to resolve potential scaling issues. The GP is configured to use Matern kernels with $\nu=0.5$. See Example-5-MC-GP-IRS-CVA.ipynb in Github for further details.

\begin{figure}[h!]
\centering

\includegraphics[width=0.7\linewidth]{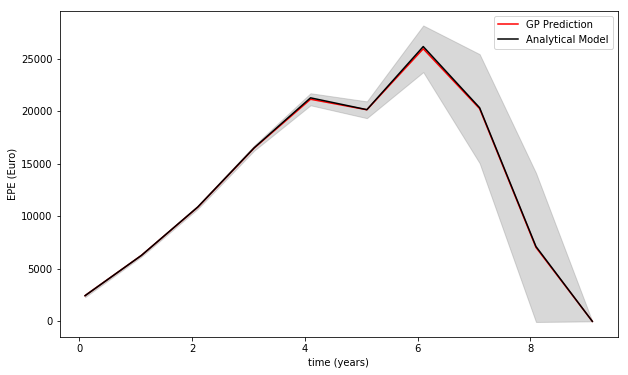} 
\caption{\textit{The EPE profile (Euros) of the IRS portfolio as evaluated using an analytic model versus the GP (with Matern Kernels). The grey band denotes the 95\% GP uncertainty band.}}  
\label{fig:EPS_0_IRS}
\end{figure}

The EPE profile (cf. Section \ref{ss:epecva}) of the IRS portfolio evaluated using the pricing formula (MC-reval) versus the MC-GP is shown in Figure \ref{fig:EPS_0_IRS}. The $\textrm{CVA}_0$ using the MC-GP or MC-reval model is given in Table \ref{tab:CVA_0_IRS}. The mean of GP model estimate $\widehat{\textrm{CVA}}_0$ is found to be within $0.25\%$ of the GP-reval estimate $\textrm{CVA}_0$. The 95\% upper and lower confidence intervals for the MC sampling are also given. 

\begin{table}
\centering
\begin{tabular}{c|ccc}
\hline
Estimator & Mean & lower C.I. & upper C.I.\\
\hline
$\textrm{CVA}_0$     &  3357.846 & 3179.715 & 3535.978\\
$\widehat{\textrm{CVA}}_0$  & 3177.103 & 3522.072 & 3571.0287\\
\hline
\end{tabular}
\label{tab:CVA_0_IRS}
\caption{\textit{The $\textrm{CVA}_0$ using the MC-reval model is compared with the mean of GP model, $\widehat{\textrm{CVA}}_0$, and is found to be within $0.25\%$ of the GP-reval estimate. The 95\% upper and lower confidence intervals for the MC sampling are also given.}} 
\end{table}


\section{Conclusion} \label{sect:conclusion}

This paper introduces a Gaussian process regression and Monte Carlo simulation  (MC-GP) approach for fast evaluation of derivative portfolios, their sensitivities, and related counterparty credit risk metrics such as the EPE and the CVA. The approach is demonstrated by estimating the CVA on a simple portfolio with numerical studies of accuracy and convergence (in terms of both the numbers of training points and of sample paths) of our MC-GP estimates. 
Once the kernels have been learned, there is no need to use expensive derivative pricing  or Greeking functions. The kernels permit a closed form approximation for the sensitivity of the portfolio to the risk factors and the approach preserves the flexibility to rebalance the portfolio. Efficient hyper-parameter optimization procedures are available. 
Moreover, the advantage is not just computational: The risk estimation approach is Bayesian---the uncertainty in a point estimate which the model hasn't seen in the training data is quantified. 
The approach is scalable 
through a divide-and-conquer approach, possibly implemented in parallel, where a different GP is used for each sub-portfolio of assets depending on the same (few) risk factors.


Compared with simpler alternatives such as splines or kernel smoothers,
GP regressions offer a metamodeling framework with a probabilistic Bayesian interpretation and a quantification of the associated numerical uncertainty. Marginal likelihood maximization yields a convenient way of setting the 
hyperparameters.
GPs can cope with noisy data but they are also interpolating in the noise-free limit.
As opposed to Chebyshev interpolation, which uses a deterministic node location imposed by the scheme (in conjunction with suitable interpolation weights, see \citep{GassGlauMair}), 
GPs  can use an arbitrary, possibly unstructured (e.g. stochastically simulated) grid of observations.

Compared with richer alternatives such as deep neural networks (DNN), 
GPs typically require much less data to train.
They also inherently provide ``differential regularization'' without the need to adopt cumbersome cross-validation techniques to tune regularization hyper-parameters, as in DNNs.
Also, despite recent Bayesian deep learning developments meant to enable deep learning in small data domains, DNNs  are still difficult to cast in a Bayesian framework.
However, unlike DNNs, a kernel view does not give any hidden representations, failing to identify the useful features for solving a particular problem. More elaborate choice of priors can be used to address this issue.


Our usage of ``uncertainty'' in this paper  refers to the GP regression error estimate. However, GPs could also potentially be used for uncertainty quantification in the sense of quantifying model risk. Model risk is, in particular, an important and widely open XVA issue, which we leave for future research. 

\appendix
\section{Multi-Factor Rates Model} \label{sect:rates}

We consider a correlation matrix $R$. 
For every currency $i$, we consider a vector of independent
Brownian motions $W^{(i)}$, where the superscript $i$ indicates that the
process is a Brownian motion in the $\mathbb{Q}^{(i)}$ world, which in turn is
where cash-flows in the currency $i$ are priced. Also, $e_i$ denotes the  vector with $i^{th}$ coordinate equal to 1 and all other components equal to 0.

The Hull-White model for short rates is given by

\[ \mathd r_i (t) = (\theta_i (t) - a_i r_i (t)) \mathd t + \sigma_i 
   \left\langle R^{\frac{1}{2}} e_i, \mathd W_t^{(i)} \right\rangle \]
with $\theta_i$ an exogenous deterministic function to be able to fit the
forward curve at time $0$.

Or equivalenty, we may write $r_i (t) = x_i (t) + \beta_i (t)$ with $\beta_i$
being a deterministic function and:
\[ \left\{ \begin{array}{ccl}
     \mathd x_i (t) & = & - a_i x_i (t) \mathd t + \sigma_i  \left\langle
     R^{\frac{1}{2}} e_i, \mathd W_t^{(i)} \right\rangle\\
     x_i (0) & = & 0
   \end{array} \right. \]
If $f_i (0, .)$ is the time-$0$ instantaneous forward curve for rate $i$,
then:
\[ \forall t \geq 0, \beta_i (t) = f_i (0, t) + \frac{\sigma_i^2}{2 a_i^2}  (1
   - e^{- a_i t})^2 \]
    
The FX rates are given by a mean-reverting process as follows. Consider, for ease of exposition to case when $n=1$, 
\[ \frac{\mathd \op{FX}_{j, i} (t)}{\op{FX}_{j, i} (t)} = (r_i (t) - r_j
   (t)) \mathd t + \alpha_{j, i} \sigma_3  \left\langle R^{\frac{1}{2}} e_3,
   \mathd W_t^{(i)} \right\rangle \]
where $\alpha_{1, 2} = - 1$ and $\alpha_{2, 1} = 1$. By comparing with the
dynamics obtained by FX inversion, one can deduce the following Brownian
motion change between the $\mathbb{Q}^{(1)}$ and the $\mathbb{Q}^{(2)}$
worlds:
\[ \mathd W_t^{(2)} = \mathd W_t^{(1)} - \sigma_3 R^{\frac{1}{2}} e_3 \mathd t
\]
or equivalently:
\[ \begin{array}{ccc}
     \left( \frac{\mathd \mathbb{Q}^{(2)}}{\mathd \mathbb{Q}^{(1)}} \right)_t
     & = & \exp \left( \int_0^t \sigma_3  \left\langle R^{\frac{1}{2}} e_3,
     \mathd W_s^{(i)} \right\rangle - \frac{1}{2}  \int_0^t \sigma_3 ^2
     \left\| R^{\frac{1}{2}} e_3 \right\|^2 \mathd s \right)\\
     & = & \frac{\op{FX}_{2, 1} (t)}{\op{FX}_{2, 1} (0)} \exp \left(
     \int_0^t (r_2 (s) - r_1 (s)) \mathd s \right)
   \end{array} \]

\bibliographystyle{chicago}
\bibliography{main}
\end{document}